\documentclass[journal]{IEEEtran}
\usepackage{cite}
\usepackage{gensymb}
\usepackage{graphicx}
\usepackage{bmpsize}
\usepackage{float}
\usepackage{amsmath}
\usepackage{amssymb}
\usepackage[draft]{hyperref}
\usepackage{setspace}
\usepackage{color}
\usepackage[caption=false]{subfig}
\usepackage{stfloats}
\usepackage{multirow}

\newtheorem{theorem}{Theorem}
\newtheorem{lemma}{Lemma}
\newtheorem{remark}{Remark}
\newtheorem{case}{Case}

\newcounter{MYtempeqncnt}
\IEEEoverridecommandlockouts

\makeatletter
\newcommand{\vast}{\bBigg@{5}}
\makeatother

\begin{document}

\title{Impact of Angular Spread in Moderately Large MIMO Systems under Pilot Contamination}

\author{Nadisanka~Rupasinghe,~Yavuz~Yap{\i}c{\i},~Jorge~Iscar,~and~\.{I}smail~G\"{u}ven\c{c}, \emph{Senior Member, IEEE}
\thanks{N.~Rupasinghe,~Y.~Yap{\i}c{\i},~and~\.{I}.~G\"{u}ven\c{c} are with the Department of Electrical and Computer Engineering, North Carolina State University, Raleigh, NC, 27606 (e-mail:~\{rprupasi,yyapici,iguvenc\}@ncsu.edu).

J.~Iscar was with Florida International University, Miami, FL, 33174  (e-mail: jisca001@fiu.edu).

This work is supported in part by the National Science Foundation under the grant number CNS-1618692.
}}%


\maketitle

\begin{abstract}
Pilot contamination is known to be one of the main bottlenecks for massive multi-input multi-output (MIMO) networks. For moderately large antenna arrays (of importance to recent/emerging deployments) and correlated MIMO, pilot contamination may not be the dominant limiting factor in certain scenarios. To the best of our knowledge, a rigorous characterization of the achievable rates and their explicit dependence on the angular spread (AS) is not available in the existing literature for moderately large antenna array regime. In this paper, considering eigen-beamforming (EBF) precoding, we derive an exact analytical expression for achievable rates in multi-cell MIMO systems under pilot contamination, and characterize the relation between the AS, array size, and respective user rates. Our analytical and simulation results reveal that the achievable rates for both the EBF and the regularized zero-forcing (RZF) precoders follow a non-monotonic behavior for increasing AS when the antenna array size is moderate. We argue that knowledge of this non-monotonic behavior can be exploited to develop effective user-cell pairing techniques.
\end{abstract}

\section{Introduction}



``Massive'' multi-input multi-output (MIMO) is a recent technology that can significantly improve the spectral/energy efficiency of future wireless networks~\cite{Marzetta14MassiveMIMO,Marzetta15MasMIMO,Marzetta16TenMyth}, and hence can help to address the exponentially growing traffic demand due to proliferation of smart devices. Interest in time-division-duplexing (TDD) massive MIMO systems has recently surged \cite{Marzetta10NonCoo,Marzetta15MasMIMO, Fernandes13Marzetta_InterCellInt, Gesbert13CooApp, Babis12Huh, Nadisanka16DirTra, Jorge16Nadisanka, Ashikhmin12Marzetta_Pilot_Cont_precoding}, due, in part, to their inherent scalability with the number of base station (BS) antennas where a single UL pilot trains the whole BS array. In particular, in TDD massive MIMO systems, the channel state information at the transmitter (CSIT) can be obtained by leveraging the channel reciprocity \cite{Debbah13MasMIMO}.
However, when the user density gets larger in a TDD massive MIMO network (e.g., as in urban areas), the scarcity of pilot resources necessitates the pilot resource reuse by the user equipments (UEs) in different cells. This results in pilot contamination, which impairs the orthogonality of the downlink transmissions from different BSs in TDD networks, diminishing the achievable aggregate capacity.

Adverse effects of pilot contamination in TDD-MIMO networks have been studied extensively in the recent literature, e.g., see the survey~\cite{PilotConSurvey} and the references therein. In particular, the pioneering papers~\cite{Marzetta10NonCoo,Marzetta11PilCon} define and investigate the pilot contamination problem over an \emph{uncorrelated} MIMO channel. In~\cite{Debbah13MasMIMO}, analytical rate expressions are derived for \emph{correlated} MIMO channels under pilot contamination. However, this analysis is done only for the \emph{asymptotic regime} considering very large antenna array sizes. In~\cite{Gesbert13CooApp}, pilot contamination is considered over a correlated MIMO channel, for finite and large antenna array regimes. Considering an asymptotic analysis, the adverse impacts of the pilot contamination are discussed to be completely eliminated with \emph{large antenna arrays}. This is achieved when the UL beams have \textit{non-overlapping} angular support, which can only be satisfied with small AS values. In a follow-up work~\cite{Gesbert16RobPilDec}, the power-domain separation of the desired and the interfering user channels is considered to overcome pilot contamination issue (also studied in~\cite{Muller14BliDecon}).

The pilot contamination effect has a strong connection with the AS of the propagation environment. Interestingly, the existing literature lacks a rigorous analysis for the explicit effect of the AS on the user rates under pilot contamination. In particular, focusing on \emph{correlated MIMO channels} and \emph{moderately large antenna array sizes} with $10\,{-}\,100$ antenna elements are important since these are common scenarios in present real world deployments. For instance, the $3$rd generation partnership project (3GPP) group is currently focusing on millimeter-wave (mmWave) transmissions~\cite{3GPP16ChanMod} which consists of correlated MIMO channels due to limited AS (which is frequency dependent). In addition, mmWave transmission is also receiving high attention for vehicular communication mainly due to the possibilities of generating highly directional beams \cite{Directional_Beam_mmWave} (without much interference), and providing high bandwidth for connected vehicles \cite{HighBW_mmWave}. The new radio (NR) techniques for the $5$th generation ($5$G) wireless communication are considering moderate array sizes even at mmWave frequencies \cite{3GPP16NR}, i.e., at $30$~GHz, 128 antenna elements (single polarized) in uniform planar array (UPA). For long term evolution (LTE) systems operating at sub-$6$~GHz frequencies \cite{3GPP3D16ChanMod} the number of antennas considered is even smaller, i.e., maximum 32 antenna elements (single polarized) in UPA \cite{3GPPTR36_897}. Further, for drone based communication networks \cite{Nadisanka16GC,NadisankaTCoM_arXiv} and moving networks (MNs) \cite{MN_Sui2015}, having a large antenna array is not practically feasible due to the availability of limited form factor. As a result, it becomes crucial to operate with moderate size antenna arrays for such vehicular communication networks.

In this study, we investigate the impact of the AS on the achievable user rates for a TDD based transmission over correlated MIMO channels. In particular, the effect of pilot contamination on achievable rates is analyzed with a special focus on the moderate antenna array size regime. The specific contributions of this work, which is a rigorous extension of~\cite{Jorge17PilConAS}, can be summarized as follows:
\begin{itemize}
\item[i.] An exact analytical expression for the achievable rate is derived considering eigen-beamforming (EBF) precoding explicitly taking in to account the impact of AS. In contrast to the earlier work in the literature~\cite{Debbah13MasMIMO,Gesbert13CooApp}, this analysis is valid for \textit{any} antenna array size, 
which is verified to match perfectly with the simulation data under various settings.
\item[ii.]
We show analytically that although large AS leads to stronger pilot contamination for the EBF precoding by impairing the interference channel orthogonality, this does not necessarily degrade the ergodic rates when the array size is \textit{moderate}. Interestingly, fluctuation of the channel power around its long-term mean reduces (similar to the so-called channel hardening effect~\cite{Hochwald04ChanHard, Tarokh09ChanHard, Larsson17ChanHard}) with the increasing AS, which in turn improves achievable rates for the EBF precoding.
\item[iii.] We show that the achievable rates of the EBF and the regularized zero-forcing (RZF) precoders exhibit a non-monotonic behavior with respect to the AS for \emph{moderate} antenna array sizes. The AS that results in a minimum/maximum  rate depends on the relative positions of UEs and their serving BSs. Hence, the potential for developing efficient user-cell pairing algorithms based on the derived rate expression is also discussed with the purpose of maximizing the network throughput.
\end{itemize}

\noindent
Table~\ref{tab:LitReview} places the specific contribution of our work in the context of the existing literature. Note that while  \cite{Gesbert13CooApp,Gesbert16RobPilDec} present simulation results with specific AS values for correlated MIMO and moderate number of antennas, analytical characterization of the achievable rates explicitly as a function of the AS is not carried out.

\begin{table}
\centering
\caption{Comparison of our work with the existing literature.}
\begin{tabular}{ | c | c | c | c | }
  \hline			
  \multirow{2}{*}{\textbf{Reference}}  &  \textbf{Number of} &  \textbf{Channel} &  {\textbf{Investigation of}} \\
  									   &  \textbf{antennas}  &   \textbf{type}   &  \textbf{Angular spread}  \\ \hline
  \cite{Marzetta10NonCoo}      & Asymptotic    	 & Uncorrelated & No \\ \hline
  \cite{Gesbert13CooApp}	   & Moderate  	 	 & Correlated 	& No \\ \hline
  \cite{Debbah13MasMIMO}       & Asymptotic  	 & Correlated 	& No \\ \hline
  \cite{Marzetta11PilCon}      & Moderate   	 & Uncorrelated & No \\ \hline
  \cite{Gesbert16RobPilDec}    & Moderate   	 & Correlated 	& No \\ \hline
  Our work      			   & Moderate	 	 & Correlated 	& Yes \\ \hline
\end{tabular}
\label{tab:LitReview}
\end{table}

The rest of the paper is organized as follows: Section~\ref{sec:system_model} introduces the system model for a multi-cell, TDD-based correlated MIMO network along with UL channel training under pilot contamination. An exact analytical expression to calculate achievable DL rates with the EBF precoding is derived for a given AS and an arbitrary array size in Section~\ref{sec:ul_dl_trans}. The individual power terms constituting the achievable rate expression are further investigated for the EBF precoding in Section~\ref{sec:individual_powers}, in order to develop insights on the explicit behavior of the ergodic rate as a function of the AS. Extensive numerical results are provided in Section~\ref{sec:numerical_results}, and finally, Section~\ref{sec:conclusion} provides some concluding remarks.

\textit{Notations:} Bold and uppercase letters represent matrices whereas bold and lowercase letters represent vectors. $\textbf{A}(m,n)$ denotes the $m$th row and $n$th column element of matrix $\textbf{A}$. $\|\cdot\|$, $|\cdot|$, $\left(\cdot\right)^{\rm T}$, $\left(\cdot\right)^{\rm H}$, $\left(\cdot\right)^{\rm \ast}$, ${\rm tr}\left(\cdot\right)$, $\otimes$, ${\rm Var\{\cdot\}}$ and $\mathbb{E}\{\cdot\}$ represent the Euclidean norm, absolute-value norm, transpose, Hermitian transpose, complex conjugation, trace of a matrix, Kronecker product, statistical variance and expectation operators, respectively. $\mathcal{CN}(\textbf{m},\textbf{C})$ denotes the complex-valued multivariate Gaussian distribution with the mean vector $\textbf{m}$ and the covariance matrix $\textbf{C}$, and ${\mathcal{U}[a,b]}$ denotes the continuous Uniform distribution over the interval ${[a,b]}$. $\textbf{I}_M$ and $\textbf{0}_M$ are the $M{\times}M$ identity matrix and zero matrix respectively, and $\delta (a,b)$ is the Kronecker delta function taking $1$ if $a\,{=}\,b$, and $0$ otherwise. $\xrightarrow{\,\text{P}\,}$ denotes the convergence in probability.

\section{System Model} \label{sec:system_model}
We consider a multi-cell scenario with $N_{\rm L}$ cells where each cell includes a single BS equipped with a uniform linear antenna array (ULA) of size $M$. In each cell, a total of $K$ UEs each with a single antenna are being served by their respective BSs under perfect time-synchronization. Since we are dealing with pilot contamination with varying AS, we assume that there is one UE in each cell that employs the same pilot sequence with other UEs in other cells during the UL channel estimation. By this way, all the users in this multi-cell layout are contributing to the pilot contamination, and the scenario where multiple UEs employ non-orthogonal pilot sequences in each cell remains a straightforward extension. Note that the perfect time-synchronization assumption is arguably the worst condition in terms of the pilot contamination as any synchronization approach will make the pilot sequences more \textit{orthogonal}, and hence reduce the pilot contamination ~\cite{Gesbert13CooApp}.
\begin{figure}
\begin{center}
\includegraphics[width=0.45\textwidth]{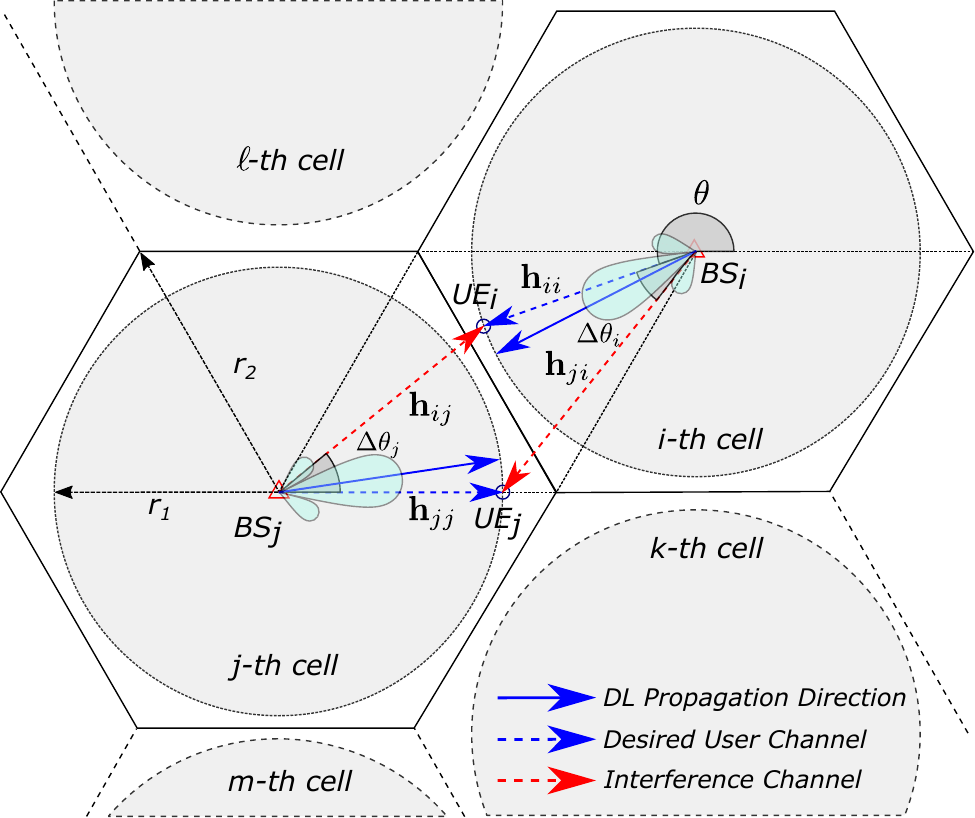}
\end{center}
\caption{The multi-cell network consisting of hexagonal cells with the side length $r_2$. All the UEs are dropped at a distance of $r_1$ from their serving BSs. The $i$th and $j$th UEs are located at $\theta$ and $0^{\circ}$ with respect to the horizontal axis measured from $i$th and $j$th BSs, respectively. $\Delta \theta_i$ denotes the angle between the directions from $i$th BS to the $i$th and $j$th UEs.}
\label{fig:layout}
\end{figure} 

In our analysis, we assume a TDD protocol consisting of subsequent UL training and DL transmission phases, where the interaction between two adjacent cells are sketched for the DL transmission in Fig.~\ref{fig:layout}. In the UL training phase, all UEs transmitting the same pilot sequence is received by all the BSs in the network. Based on the received pilot sequence, each BS first estimates the channel to its desired UE and then computes the precoding vector based on this estimated channel. During the DL transmission phase, each BS transmits data to its desired UE employing the DL precoding. Note that, due to the pilot contamination, the precoding vector is not aligning well with its desired user channel. Hence, as shown in Fig.~\ref{fig:layout} DL propagation direction is not the same as the desired user channel direction.

The UL channel between the $i$th UE and the $j$th BS $\textbf{h}_{ij}$ is
\begin{equation}
\label{eqn:channel_defn}
\begin{aligned}
\textbf{h}_{ij} = \frac{1}{\sqrt{N_\mathrm{P}}} \sum\limits_{p=1}^{N_\mathrm{P}} \alpha_{ij,p} \, \textbf{a}\left( \phi_{ij, p}\right),
\end{aligned}
\end{equation}
where $N_\mathrm{P}$ is the number multi-path components (MPCs), $\alpha_{ij,p}$ is the complex path attenuation, $\phi_{ij, p}$ is the angle of arrival (AoA) of the $p$-th MPC, and $\textbf{a}\left( \phi_{ij, p}\right)$ is the steering vector given as
\begin{equation} \small
\label{eqn:steering_vector}
\begin{aligned}
\textbf{a}\left( \phi_{ij, p}\right) = \left[ 1 \;\; e^{-j2\pi \frac{D}{\lambda}\cos\left( \phi_{ij,p}\right) } \; \dots \; e^{-j2\pi \frac{D}{\lambda}\left( M-1\right)\cos\left( \phi_{ij, p}\right) } \right]^{\rm T}  \! ,
\end{aligned} \normalsize
\end{equation}
where $D$ is the element spacing in the ULA, and $\lambda$ is the wavelength.
The complex path attenuation $\alpha_{ij, p}$ and the AoA $\phi_{ij, p}$ are assumed to be uncorrelated over any of their indices, and with each other. In particular, $\alpha_{ij, p}$ is circularly symmetric complex Gaussian with $\alpha_{ij, p}\,{\sim}\,\mathcal{CN}\left(0,\beta_{ij}\right)$, and the variance $\beta_{ij}\,{=}\,\zeta/d_{ij}^{\gamma}$ captures the effect of the large-scale path loss, where $d_{ij}$ is the distance between $i$th UE and $j$th BS, $\gamma$ is the path loss exponent, and $\zeta$ is the normalization parameter to achieve a given signal-to-noise ratio (SNR) at the BS \cite{Gesbert13CooApp}. We consider uniform distribution for the AoA with $\phi_{ij, p}\,{\sim}\,\mathcal{U}\left[\bar{\phi}_{ij}-\Delta,\bar{\phi}_{ij}+\Delta\right]$, where $\bar{\phi}_{ij}$ is the line-of-sight (LoS) angle between $i$th UE and the $j$th BS, and $\Delta$ is the AS. With the channel model in \eqref{eqn:channel_defn}, next, we study how to achieve UL training and channel estimation.
\subsection{UL Training and Channel Estimation with Correlated MIMO Channels}\label{sec:ul_trans} 
In the UL training phase, the UEs transmit the common pilot sequence of size $\tau$ denoted by $\textbf{s}\,{=}\, [ s_1 \, s_2 \, \dots s_\tau ]^{\rm T}$, where each pilot symbol is chosen in an independent and identically distributed (iid) fashion from a discrete alphabet $\mathcal{A}_{\,\rm UL}$ consisting of unity norm entries. The $M{\times}\tau$ matrix of the received symbols at the $j$th BS is given as \begin{align}\label{eqn:UL_trans_matrix}
\textbf{Y}_{j}^{\,\textrm{UL}} = \sum \limits_{i=1}^{N_\mathrm{L}} \textbf{h}_{ij} \textbf{s}^{\textrm{T}}+\textbf{N}_{j}~,
\end{align} where $\textbf{N}_{j}$ is a $M{\times}\tau$ noise matrix consisting of circularly symmetric complex Gaussian entries with $\mathcal{CN}\left(0,\sigma^2\right)$. In an equivalent vector representation, \eqref{eqn:UL_trans_matrix} is given as
\begin{equation} \label{eqn:UL_trans_vector}
\textbf{y}_{j}^{\,\textrm{UL}}=\textbf{S} \sum \limits_{i=1}^{N_\mathrm{L}} \textbf{h}_{ij} +\textbf{n}_{j}~,
\end{equation}
where $M\tau{\times}1$ vectors $\textbf{y}_{j}^{\,\textrm{UL}}$ and $\textbf{n}_{j}$ are obtained by stacking all columns of $\textbf{Y}_{j}^{\,\textrm{UL}}$ and $\textbf{N}_{j}$, respectively, and $\textbf{S}\,{=}\,\textbf{s} \otimes \textbf{I}_M$ is the training matrix of size $M\tau{\times}M$ satisfying $\textbf{S}^H\textbf{S}\,{=}\,\tau \textbf{I}_M$. Following the convention of~\cite{Marzetta10NonCoo}, the SNR is defined for this particular phase to be $1{/}\sigma^2$.

At the $j$th BS, the channel to the $i$th UE can be estimated using linear minimum mean square error (LMMSE) criterion as follows~\cite{Gesbert13CooApp}
\begin{equation} \label{eqn:mmse_chan_est}
\hat{\textbf{h}}_{ij} = \tilde{\textbf{R}}_{ij}\textbf{S}^{\rm H}\textbf{y}_{j}^{\,\textrm{UL}}~,
\end{equation}
where $\tilde{\textbf{R}}_{ij}$ is the pilot-independent estimation filter given as
\begin{equation} \label{eqn:mmse_filter}
\tilde{\textbf{R}}_{ij} = \textbf{R}_{ij} \left(\sigma^2 \textbf{I}_{M}+\tau \sum \limits_{\ell=1}^{N_\mathrm{L}}\textbf{R}_{\ell j}  \right)^{{-}1} \!\! .
\end{equation}
The covariance matrix $\textbf{R}_{ij}\,{=}\,\mathbb{E}\left\lbrace \textbf{h}_{ij} \textbf{h}_{ij}^{\rm H}\right\rbrace$ in~\eqref{eqn:mmse_filter} is defined element-wise as follows \vspace{-1.5em}

\small
\begin{align} \label{eqn:cov_matrix}
&\mathbf{R}_{ij}(m,n) = \beta_{ij} \mathbf{R}_{ij}^{\phi}(m,n)
\\ \nonumber
&= \beta_{ij} \int_{0}^{2\pi} \exp \left( {-}j 2\pi (m-n)\frac{D}{\lambda} \cos(\phi_{ij}) \right) p_{\phi}(\phi){\rm d} \phi~,
\end{align} \normalsize
where $p_{\phi}(\phi)$ is the probability distribution function (pdf) of the AoA, and $\mathbf{R}_{ij}^{\phi}\,{=}\,\mathbb{E}\left\lbrace \textbf{a}\left(\phi_{ij}\right) \textbf{a}^{\rm H}\left( \phi_{ij}\right) \right\rbrace$ is the angular covariance matrix of the steering vector. Employing \eqref{eqn:mmse_filter} and \eqref{eqn:cov_matrix}, the resulting covariance matrix of the channel estimate in \eqref{eqn:mmse_chan_est} is given as
\begin{equation}\label{eqn:cov_matrix_mmse}
\hat{\textbf{R}}_{ij}\,{=}\,\mathbb{E} \big\{ \hat{\textbf{h}}_{ij} \hat{\textbf{h}}_{ij}^{\rm H} \big\}\,{=}\,\tau \tilde{\textbf{R}}_{ij} \textbf{R}_{ij}\,,
\end{equation}
where we present the detailed derivation steps for covariance matrices in Appendix~\ref{app:cov_matrix}.

\section{Achievable DL Rates for Correlated MIMO Channels with Moderate Antenna Array Sizes}\label{sec:ul_dl_trans}

In this section, we study the achievable DL rates for correlated MIMO channels specifically considering the moderate size antenna array regime. In particular, we derive an exact analytical expression to calculate achievable DL ergodic rates with eigen-beamforming (EBF) precoding under pilot contamination. This rate expression is applicable to \emph{any} antenna array size, unlike the case in \cite{Debbah13MasMIMO} where \emph{asymptotic} antenna array regime is taken in to consideration.

During the DL data transmission, each BS employs the channel estimate obtained in the UL training phase as discussed in Section~\ref{sec:ul_trans} to compute the precoding vector for its own UE relying on the perfect reciprocity of the UL and the DL channels in the TDD protocol~\cite{Marzetta10NonCoo}. The received signal at the $j$th UE can therefore be given as
\begin{equation} \label{eqn:DL_trans}
y_{j}^{\rm DL} = \sqrt{\eta_{j}}\textbf{h}_{jj}^{\rm H}\textbf{w}_{j} q_{j}  + \sum \limits_{i=1; i \neq j}^{N_\mathrm{L}} \sqrt{\eta_{i}}\textbf{h}_{ji}^{\rm H}\textbf{w}_{i} q_{i} + n_{j},
\end{equation}
where $\textbf{w}_{i}$ is the $M{\times}1$ precoding vector of the $i$th BS for its own user, $\eta_{i}\,{=}\,\left[ \mathbb{E} \left\lbrace {\rm tr} \left[ \textbf{w}_{i} \textbf{w}_{i}^{\rm H} \right] \right\rbrace \right]^{{-}1}$ normalizes the average transmit power of the $i$th BS to achieve the same SNR in the UL training phase~\cite{Debbah13MasMIMO}, $q_j$ is the unit-energy data symbol transmitted from $j$th BS to its own UE and chosen from a discrete alphabet $\mathcal{A}_{\,\rm DL}$ in an iid fashion, and $n_j$ is the circularly symmetric complex Gaussian noise with $\mathcal{CN}\left(0,\sigma^2\right)$. The beamforming strategy is assumed to be either the EBF (also known as conjugate beamforming) or the regularized zero-forcing (RZF)~\cite{Debbah13MasMIMO}, and is given at the $j$th BS as follows
\begin{align}
\textbf{w}^{\rm EBF}_j &= \hat{\textbf{h}}_{jj}\,, &\text{(EBF Precoder)} \label{eqn:ebf}\\
\textbf{w}_j^{\rm RZF} &= \left( \hat{\mathbf{h}}_{jj} \hat{\mathbf{h}}_{jj}^{\rm H} + \sigma^2 \textbf{I}_{M} \right)^{-1}\hat{\mathbf{h}}_{jj}\,. &\text{(RZF Precoder)} \label{eqn:rzf}
\end{align} In the following, the impact of AS on the achievable rates is investigated under both of these beamforming strategies, with a rigorous analytical rate derivation for the EBF precoding.

\subsection{Achievable DL Rates with Precoding} \label{sec:rates}

\begin{figure*}[!t]
\setcounter{MYtempeqncnt}{\value{equation}}
\setcounter{equation}{11}
\begin{equation} \label{eqn:ergodic_rate}
R_{j} = \log_2 \vast( 1 + \frac{\overbrace{\eta_{j} \left| \mathbb{E} \left\lbrace \textbf{h}_{jj}^{\rm H} \textbf{w}_j \right\rbrace \right|^2}^{\textrm{Desired signal power}}}{\sigma^2\,{+}\,\underbrace{\eta_{j}{\rm Var}\left\lbrace \textbf{h}_{jj}^{\rm H}\textbf{w}_j\right\rbrace}_{\textrm{Self-interference}}\,{+}\,\underbrace{\sum_{i=1; i\neq j}^{N_\mathrm{L}}\eta_{i} \mathbb{E} \left\lbrace \left| \textbf{h}_{ji}^{\rm H} \textbf{w}_i \right|^2 \right\rbrace }_{\textrm{Intercell interference}}} \vast)
\end{equation}
\setcounter{equation}{12}
\hrulefill
\end{figure*}

We now study achievable DL rates as a function of the AS over the underlying correlated MIMO channel with the EBF and the RZF precoding. In particular, we provide an exact analytical expression to calculate achievable rates with EBF precoding. By assuming UEs have just the knowledge of long-term statistics of the effective channel and not the instantaneous CSI, the ergodic rate as given in \eqref{eqn:ergodic_rate} is achievable at the $j$th UE~\cite{Marzetta11PilCon}. In that, $\eta_{j}\left|\mathbb{E}\left\lbrace \textbf{h}_{jj}^{\rm H} \textbf{w}_j \right\rbrace\right|^2$ captures the desired signal power, $\eta_{j}{\rm Var}\left\lbrace\textbf{h}_{jj}^{\rm H} \textbf{w}_j \right\rbrace$ is interpreted as the \textit{self-interference} and arises from the lack of information on the instantaneous channel at the UE, and $\sum_{i=1; i\neq j}^{N_\mathrm{L}} \eta_{i}\mathbb{E}\left\lbrace \left|\textbf{h}_{ji}^{\rm H} \textbf{w}_i \right|^2 \right\rbrace$ is the intercell interference since it represents the interference from the other BS signals. Here, the power normalization factor is given by $\eta_{j} = \left[ \mathbb{E} \big\{ \hat{\textbf{h}}_{jj}^{\rm H} \hat{\textbf{h}}_{jj} \big\} \right]^{-1}$.

The rate approximation in \eqref{eqn:ergodic_rate} is arguably conservative, as discussed in~\cite{Nadisanka16DirTra}, and can be interpreted as ``self-interference limited'' rate since the self-interference term dominates at high SNR regime for finite ULA sizes. However, since our focus in this study is to evaluate the impact of the AS on the correlated MIMO channels at fixed SNR, the rate approximation in \eqref{eqn:ergodic_rate} is used confidently. It is worth noting that, the achievable rates can also be evaluated by considering the alternative expression suggested in~\cite[Eqn.~(32)]{Nadisanka16DirTra} using the first and the second order moments of the effective channel derive subsequently.

In the following theorem, considering that the EBF precoder in~\eqref{eqn:ebf} is used in the DL transmission, we derive analytical expressions of the first and the second order moments for the effective channel in order to be able to calculate the achievable rate in~\eqref{eqn:ergodic_rate}.

\smallskip

\begin{theorem}\label{the:achievable_rates}
Assuming that LMMSE channel estimation is used in the UL training, and that  EBF precoding as in~\eqref{eqn:ebf} is used prior to DL data transmission, the first order moment of the effective channel is given as
\begin{align}\label{eqn:1st_moment}
\mathbb{E} \left\lbrace \textbf{h}_{jj}^{\rm H} \textbf{w}_j  \right\rbrace &= {\rm tr} \left\lbrace  \hat{\textbf{R}}_{jj}  \right\rbrace\,.
\end{align}


\begin{figure*}[!t]
\setcounter{MYtempeqncnt}{\value{equation}}
\setcounter{equation}{13}
\begin{align}
\mathbb{E} \left\lbrace \left| \textbf{h}_{ji}^{\rm H} \textbf{w}_i \right|^2 \right\rbrace &= \tau^2 \sum\limits_{m=1}^{M} \sum\limits_{n=1}^{M} \sum\limits_{m'=1}^{M} \sum\limits_{n'=1}^{M} \tilde{\textbf{R}}_{ii}(m,n) \, \tilde{\textbf{R}}_{ii}^{\rm H}(m',n') \;
\left[ {\rm E}_{\phi}(m,n,m',n') + \sum\limits_{\substack{k=1; \, k{\neq}j}}^{N_\mathrm{L}} \!\!\!\textbf{R}_{ji}(n',m) \textbf{R}_{ki}(n,m') \right] \nonumber\\
& \qquad + \sigma^2 {\rm tr} \left\lbrace \textbf{S} \tilde{\textbf{R}}_{ii}^{\rm H}\textbf{R}_{ji} \tilde{\textbf{R}}_{ii} \textbf{S}^{\rm H} \right\rbrace , \label{eqn:2nd_moment}
\end{align}
\setcounter{equation}{14}
\vspace{-0.1in}
\begin{equation} \label{eqn:2nd_moment_phi}
{\rm E}_{\phi}(m,n,m',n')\,=\, \dfrac{\beta_{ji}^2}{{N_\mathrm{P}}} \Big[ 2 \,{\rm E}_{ji} (n{-}m{+}n'{-}m') + (N_\mathrm{P}{-}1) \Big( {\rm E}_{ji}(n{-}m){\rm E}_{ji} (n'{-}m') + {\rm E}_{ji}(n'-m){\rm E}_{ji}(n-m') \Big) \Big] \,,
\end{equation}
\setcounter{equation}{15}
\hrulefill
\end{figure*}

Likewise, the second order moment, $\mathbb{E} \left\lbrace \left| \textbf{h}_{ji}^{\rm H} \textbf{w}_i \right|^2 \right\rbrace$ can be given as in \eqref{eqn:2nd_moment} where ${\rm E}_{\phi}(m,n,m',n')$ is defined as in \eqref{eqn:2nd_moment_phi} with ${\rm E}_{ji}(m)\,{=}\,\mathbf{R}_{ji}^{\phi}(n{+}m,n)$ for any $n\,{\leq}\,M{-}m$.
\end{theorem}

\smallskip

\begin{IEEEproof}
See Appendix~\ref{app:achievable_rates}.
\end{IEEEproof}

\smallskip

Note that once \eqref{eqn:1st_moment} and \eqref{eqn:2nd_moment} are computed, the signal and the intercell interference terms in \eqref{eqn:ergodic_rate} are readily available by employing $\eta_{j}\,{=}\left[ {\rm tr} \big\{ \hat{\textbf{R}}_{jj} \big\} \right]^{{-}1}\!\!$, and the self-interference is given by
\begin{align}\label{eqn:variance}
{\rm Var} \left\lbrace \textbf{h}_{jj}^{\rm H} \textbf{w}_j \right\rbrace = \mathbb{E} \left\lbrace \left| \textbf{h}_{jj}^{\rm H} \textbf{w}_j \right|^2 \right\rbrace - \left( \mathbb{E} \left\lbrace \textbf{h}_{jj}^{\rm H} \textbf{w}_j \right\rbrace  \right)^2\,.
\end{align}

\section{Impact of AS on Desired and Interference Signal Power Terms}\label{sec:individual_powers}

In this section, we study in detail the explicit impact of AS on the \emph{desired signal power} (Section~\ref{sec:covariance_diagonal}), the \emph{intercell interference} (Section~\ref{sec:channel_orthogonal}), and the \emph{self-interference} (Section~\ref{sec:channel_fluctuation}) terms in \eqref{eqn:ergodic_rate} in relation to pilot contamination effect from statistical and geometrical perspectives. We also draw useful insights about their behavior for varying AS considering different array sizes. It is worth remarking that, any variation in the achievable rates with varying AS is due to collective contribution from all these three terms, and taking any of them only individually into account may be misleading when evaluating the overall rate results presented in Section~\ref{sec:numerical_results}.

\subsection{Effect of Covariance Matrix Diagonalization on Desired Signal Power}\label{sec:covariance_diagonal}
\begin{figure}[!h]
\centering
\captionsetup[subfigure]{oneside,margin={0.5cm,0.5cm}}
\subfloat[$|\textbf{R}_{jj}^{\phi}(m,n)|$ versus AS for various $|m{-}n|$ values.]{\includegraphics[width=0.25\textwidth]{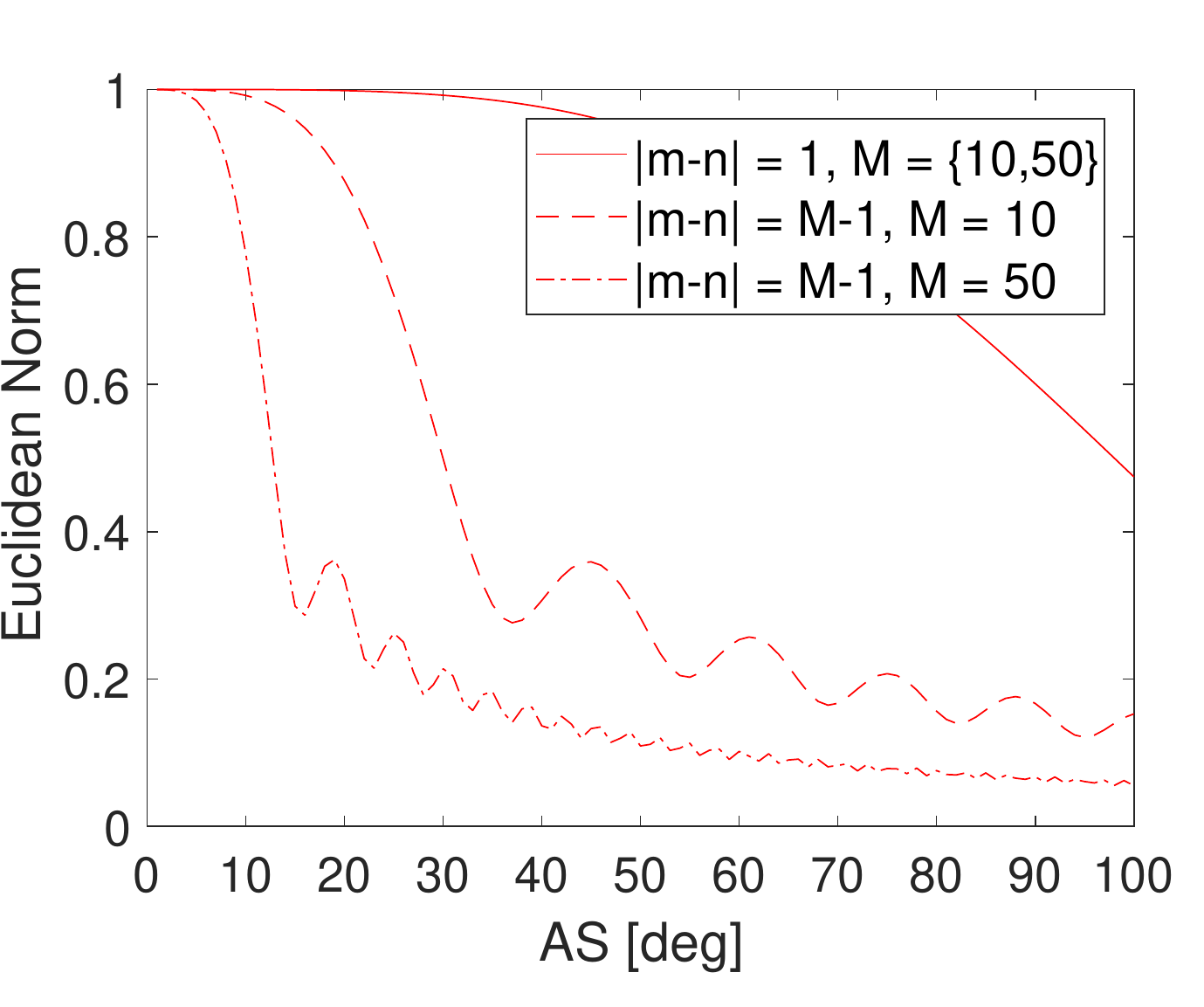}
\label{fig:r_ii}}
\subfloat[$|\textbf{R}_{ij}^{\phi}(m,n)|$ versus AS for various $|m{-}n|$ values.]{\includegraphics[width=0.25\textwidth]{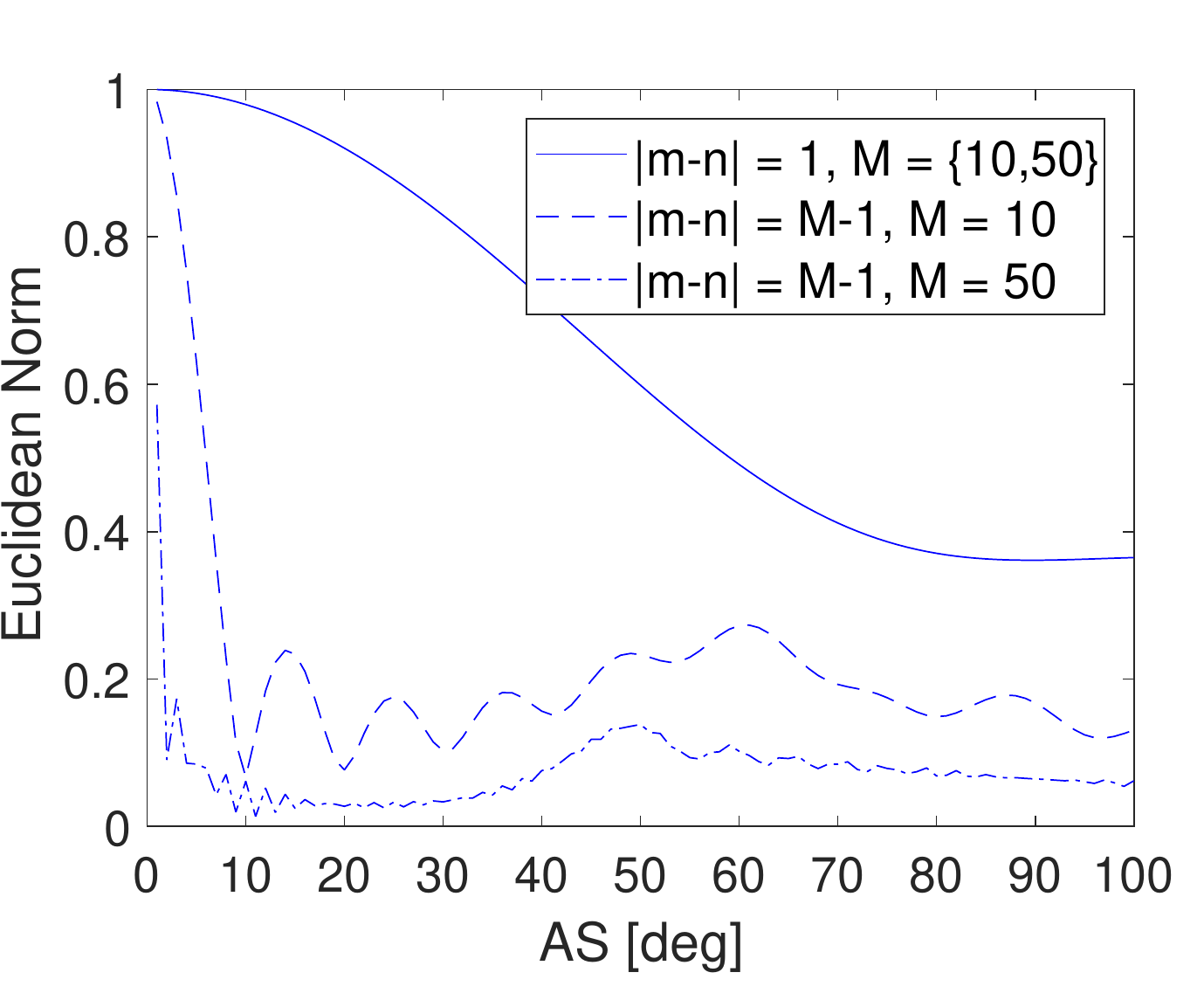}
\label{fig:r_ji}}
\vspace{0.1in}
\caption{Norm of the entries of the angular covariance matrices $\textbf{R}_{jj}^{\phi}$ and $\textbf{R}_{ij}^{\phi}$ for $M\,{=}\,\{10,50\}$.}
\label{fig:r}
\end{figure}
Let us consider the structure of the angular covariance matrices $\textbf{R}_{jj}^{\phi}$ and $\textbf{R}_{ij}^{\phi}$ in \eqref{eqn:cov_matrix} with varying AS and antenna array size. Under the assumptions in Section~\ref{sec:system_model}, the main diagonal entries of these covariance matrices are all $1$, and the off-diagonal entries representing the angular correlation have non-zero norms smaller than $1$. In Fig.~\ref{fig:r}, the Euclidean norms of the $(m,n)$th off-diagonal entries having the minimum and the maximum absolute separation of $|m{-}n|\,{=}\,1$ and $|m{-}n|\,{=}\,M{-}1$, respectively, are depicted along with the increasing AS for $M\,{=}\,\{10, \ 50\}$ and considering $\Delta\theta_j\,{=}\,40^{\circ}$ in Fig.~\ref{fig:layout}. We observe that each of these covariance matrices gets more diagonalized (the magnitude of the off-diagonal entries decreases) when the array size $M$ or the AS increases. The diagonalization rate increases with $M$ since both covariance matrices get diagonalized much faster for larger $M$ values. Note that any BS in the multi-cell network will receive signals from a wide range of AoAs as the AS increases, which is similar to the uncorrelated rich-scattering environment where the possible AoAs span $[0,2\pi)$ angle support.

The signal power variation with respect to the AS can be assessed through the diagonalization characteristic of the covariance matrices. Employing $\eta_{j}\,{=}\,\big[ {\rm tr} \big\{ \hat{\textbf{R}}_{jj}\big\} \big]^{{-}1}$ and the first order moment given in \eqref{eqn:1st_moment}, the signal power can be expressed as $\eta_{j}\left|\mathbb{E}\left\lbrace \textbf{h}_{jj}^{\rm H} \textbf{w}_j \right\rbrace\right|^2\,{=}\,{\rm tr} \big\{ \hat{\textbf{R}}_{jj}\big\}$, which can be expressed more elaborately as follows
\begin{align} \label{eqn:signal_power}
{\rm tr} \big\{ \tau\tilde{\textbf{R}}_{jj}\textbf{R}_{jj}\big\} &= \tau\sum\limits_{m}\tilde{\textbf{R}}_{jj}(m,m)\textbf{R}_{jj}(m,m) \nonumber \\
&+ \tau \mathop{\sum\limits_{m}\sum\limits_{n}}_{n \neq m}\tilde{\textbf{R}}_{jj}(m,n)\textbf{R}_{jj}(n,m) .
\end{align}
Since the estimation matrix $\tilde{\textbf{R}}_{jj}$ in \eqref{eqn:mmse_filter} becomes diagonal for larger AS values (similar to $\textbf{R}_{ij}$ and $\textbf{R}_{jj}$), and that both the terms at the right hand side of \eqref{eqn:signal_power} are real and positive, the second summation in \eqref{eqn:signal_power} decreases when the AS increases. This leads the covariance matrix to become more diagonal. As a result, the \emph{\textbf{signal power in \eqref{eqn:ergodic_rate} decreases with increasing AS}} through the diagonalization of the covariance matrices.

This behavior of the signal power with increasing AS can be intuitively interpreted as follows. As we will discuss in Section~\ref{sec:channel_orthogonal}, when the AS increases, the orthogonality between the desired user precoder $\textbf{w}_{j}\,{=}\,\hat{\textbf{h}}_{jj}$ and the interfering user channel $\textbf{h}_{ij}$ gets impaired along with more powerful pilot contamination. Hence, $\textbf{w}_{j}$ does not exactly align with the $j$th user channel direction $\textbf{h}_{jj}$, any more. This geometrical misalignment accordingly results in transmit power leakage from $j$th BS to some undesired directions (other than the $j$th user direction) during the DL data transmission, which in turn leads to signal power loss at the $j$th user.

\subsection{Geometrical Interpretation of Intercell Interference Power}\label{sec:channel_orthogonal}
In the DL data transmission, the pilot contamination shows its adverse effect by impairing the orthogonality between the desired and the interfering user channels, which is basically captured by the intercell interference term $\sum \limits_{i=1; i \neq j}^{N_\mathrm{L}} \sqrt{\eta_{i}}\textbf{h}_{ji}^{\rm H}\textbf{w}_{i} q_{i}$ in \eqref{eqn:DL_trans}. From a geometrical perspective, the intercell interference involves the inner product between each interference channel $\textbf{h}_{ji}$ for $j\,{\neq}\,i$, and the precoder $\textbf{w}_{i}$, which is a function of the estimate of the desired user channel $\textbf{h}_{ii}$. One way to examine how the pilot contamination impairs the orthogonality, and hence amplify the intercell interference power with varying AS is through a geometric interpretation. This can be done by analyzing the pdf of the random angle $\varphi_{ij}$ between $\textbf{h}_{ji}$ and $\textbf{w}_{i}$, where we leave the actual numerical evaluation to Section~\ref{sec:numerical_results}. Note that, if the channels were perfectly known and spatially uncorrelated, the desired pdf would be given analytically as $f_{\varphi_{ij}}(\varphi)\,{=}\,2\left(N{-}1\right)\left(\sin\varphi\right)^{2N{-}3}\cos\varphi$~\cite{Loyka2004PerAnVB}.

\begin{figure}[!h]
\centering
\subfloat[The distribution of $\varphi_{ij}$ for $M\,{=}\,10$.]{\includegraphics[width=0.45\textwidth]{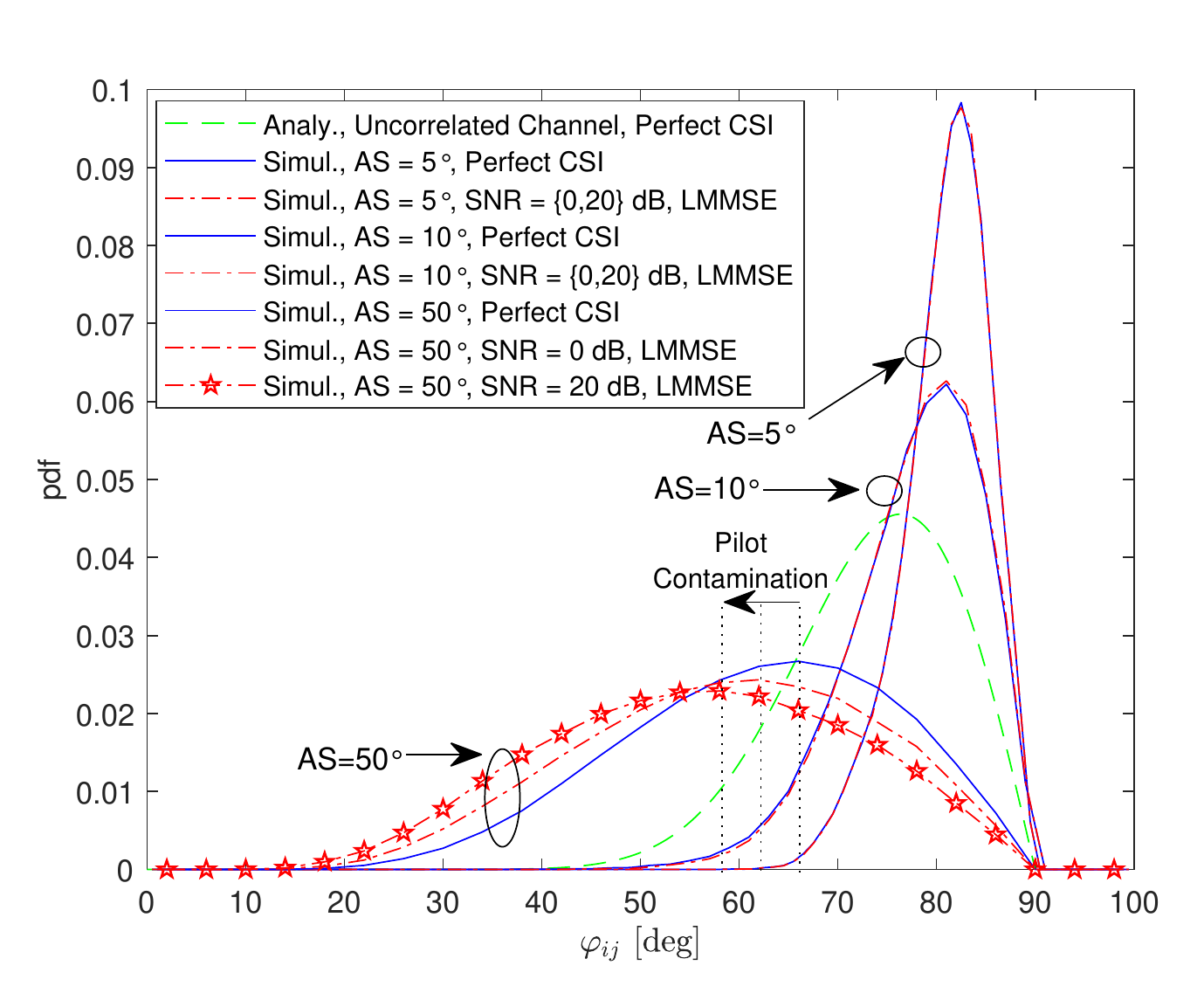}
\label{fig:pdf_M_10}}\\
\subfloat[The distribution of $\varphi_{ij}$ for $M\,{=}\,50$.]{\includegraphics[width=0.45\textwidth]{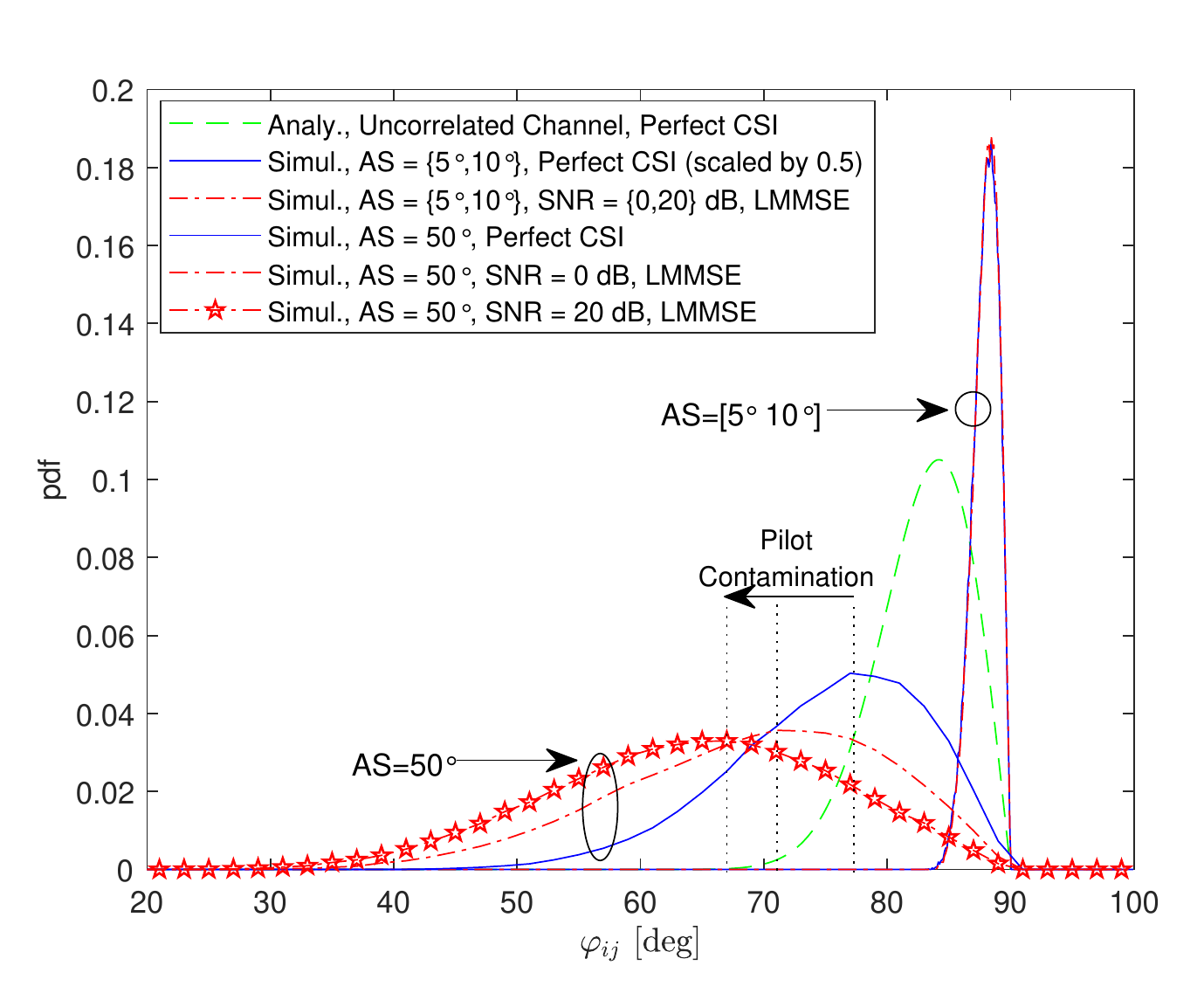}
\label{fig:pdf_M_50}}
\vspace{0.1in}
\caption{The pdf of the angle $\varphi_{ij}$ between $\textbf{h}_{ji}$ and $\textbf{w}_{i}$ for $M\,{=}\,\{10,50\}$ and $\text{SNR}\,{=}\,\{0,20\}\text{ dB}$.}
\label{fig:pdf}
\end{figure}

To study the impact of AS on intercell interference, we consider an example scenario with the representative setting of Fig.~\ref{fig:layout}. In that, we assume $\theta\,{=}\,200^{\circ}$, quadrature phase-shift keying (QPSK) symbols in the UL training phase with the sequence length $\tau\,{=}\,1$, and the path-loss exponent $\zeta\,{=}\,2$. In Fig.~\ref{fig:pdf}, we depict the pdf of the random angle $\varphi_{ij}$ for $M\,{=}\,\{10,50\}$ and $\text{SNR}\,{=}\,\{0,20\}\text{ dB}$. We observe that the desired and the interfering user channels are sufficiently orthogonal for small AS values since $\varphi_{ij}$ takes values close to $90^{\circ}$ with high probability for perfect CSI, and the resulting orthogonality can even be stronger than the uncorrelated channel case. This geometrical interpretation agrees with~\cite{Gesbert13CooApp} in the sense that the training beams of different UEs do not overlap in the UL transmission when the AS is sufficiently small making the scenario free from any pilot contamination effect. As a result, the random angle between the precoder $\textbf{w}_{i}$ and the interfering user channel $\textbf{h}_{ji}$ has the same pdf for the perfect CSI ($\textbf{w}_{i}\,{=}\,\textbf{h}_{ii}$) and the channel estimation ($\textbf{w}_{i}\,{=}\,\hat{\textbf{h}}_{ii}$) scenarios when the beams are separated sufficiently, or equivalently the AS is small enough. When the AS starts to increase, spatial correlation between the desired and the interfering user channels becomes stronger since the overlap between AoA domains associated with desired and interfering user channels becomes larger. As a result, \textbf{\emph{the desired orthogonality inherently gets impaired for larger AS values, even for the perfect CSI case}}.

When the desired user channel is being estimated, this orthogonality gets hurt even more because of the pilot contamination effect. This can be observed in Fig.~\ref{fig:pdf} from the deviation of the pdf of $\varphi_{ij}$ associated with the channel estimation scenario, to the left side (toward $0^{\circ}$) with respect to perfect CSI scenario when $\textrm{AS}\, {=}\,50^{\circ}$. The orthogonality gets impaired further when the SNR increases since larger SNR in each cell implies more interference power transferred to other cells. Finally, comparing Fig.~\ref{fig:pdf}\subref{fig:pdf_M_10} and Fig.~\ref{fig:pdf}\subref{fig:pdf_M_50} we observe that, for a given AS the precoder $\textbf{w}_{i}$ and the interfering channels are getting more orthogonal with increasing antenna array size, which is one of the main goals of massive MIMO in the context of the intercell interference rejection~\cite{Marzetta15MasMIMO}.

\subsection{Effect of Channel Power Fluctuation on Self-Interference}\label{sec:channel_fluctuation}
The rate bound given in \eqref{eqn:ergodic_rate} is discussed to be achievable in~\cite{Marzetta11PilCon} assuming that the UEs in the network do not know their instantaneous channels, but rather they only know the respective long-term means. This lack of information on the exact instantaneous channel is captured by the self-interference term $\eta_{j}{\rm Var}\left\lbrace\textbf{h}_{jj}^{\rm H} \textbf{w}_j \right\rbrace$ in \eqref{eqn:ergodic_rate}. This term actually represents the power of the deviation between the instantaneous channel and the long-term mean, given equivalently as $\eta_{j}\mathbb{E}\left\lbrace \left|\textbf{h}_{jj}^{\rm H} \textbf{w}_j{-}\mathbb{E}\left\lbrace \textbf{h}_{jj}^{\rm H} \textbf{w}_j \right\rbrace \right|^2\right\rbrace$. Assuming EBF precoding in \eqref{eqn:ebf} with perfect CSI, this term becomes equivalent to the variance of the channel power. We therefore note that as the fluctuation of the channel square-norm around the long-term mean decreases, which is similar to the phenomenon known as the \textit{channel hardening}~\cite{Hochwald04ChanHard, Tarokh09ChanHard, Larsson17ChanHard}, the self-interference term should decrease accordingly. In the following, this fluctuation and hence the \emph{\textbf{self-interference is shown to decrease monotonically when the AS increases}}, for the EBF precoding.

\smallskip

\begin{lemma}\label{lem:hardening}
The channel considered in~\eqref{eqn:channel_defn} hardens, such that $\left\| \textbf{h}_{ij} \right\|^2\!{/}\,\mathbb{E}\left\lbrace \left\| \textbf{h}_{ij} \right\|^2 \right\rbrace{\stackrel{\;\text{P}\;\;}{\rightarrow}}\,1$ as $M{\rightarrow}\infty$, if we have $\mathcal{M}_{ij}\,{\rightarrow}\,0$ as $M{\rightarrow}\,\infty$, where $\mathcal{M}_{ij}$ is the \textit{hardening measure} given as
\begin{align}\label{eqn:harden_measure}
\mathcal{M}_{ij} = \frac{1}{N_\mathrm{P}} + \frac{N_\mathrm{P}-1}{M N_\mathrm{P}} \left( 1 + \frac{2}{M} \sum\limits_{m=1}^{M-1} (M-m) \left| {\rm E}_{ij}(m) \right|^2 \right),
\end{align}
with ${\rm E}_{ij}(m)\,{=}\,\mathbf{R}_{ij}^{\phi}(n{+}m,n)$ for any $n\,{\leq}\,M{-}m$.
\end{lemma}

\smallskip

\begin{IEEEproof}
See the derivation in Appendix~\ref{app:harden_measure} as an extension of~\cite{Larsson17ChanHard} which considers uncorrelated MIMO channels.
\end{IEEEproof}

\smallskip

\begin{figure}[!t]
\begin{center}
\includegraphics[width=0.45\textwidth]{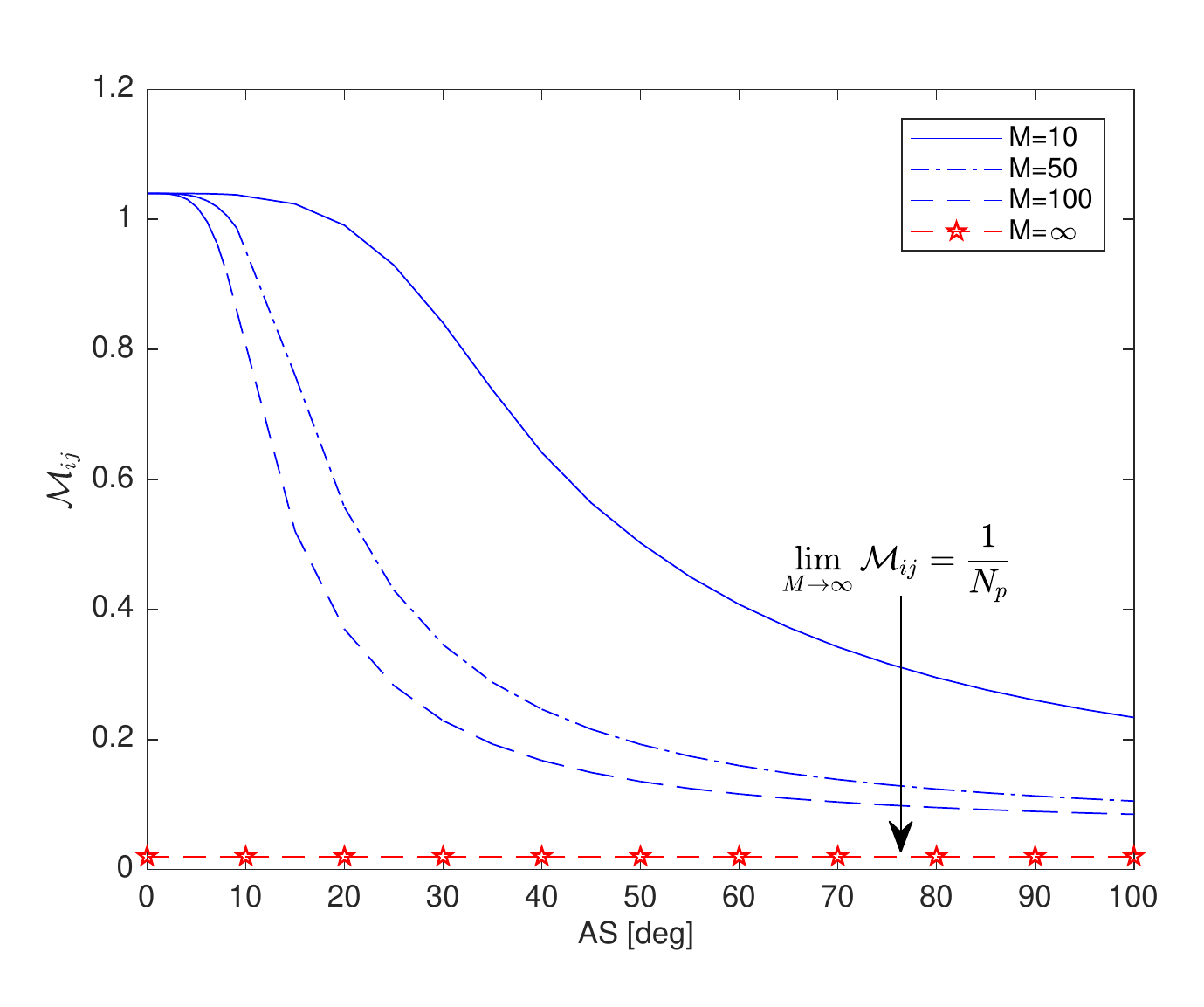}
\end{center}
\caption{The channel hardening measure $\mathcal{M}_{ij}$ for $M\,{=}\,\{10,50,100,\infty\}$ and $N_\mathrm{P}\,{=}50$.}
\label{fig:hardening}
\end{figure}

We observe that the desired convergence $\mathcal{M}_{ij}\,{\rightarrow}\,0$ is satisfied only if the number of paths $N_\mathrm{P}$ is sufficiently large. Even when the number of paths $N_\mathrm{P}$ or the antenna array size $M$ have moderate values in contrast to asymptotic approximations, the behavior of the self-interference power can still be assessed from \eqref{eqn:harden_measure}. In Fig.~\ref{fig:hardening}, we depict $\mathcal{M}_{ij}$ along with AS for various $M$ under the assumption that $N_\mathrm{P}\,{=}50$ and $\phi_{ij}\,{\sim}\,\mathcal{U}\left[-\Delta,+\Delta\right]$ with $\Delta$ representing the AS. We observe that \emph{\textbf{$\mathcal{M}_{ij}$ decreases monotonically for increasing AS for all cases, and gets even smaller values as the array size $M$ increases}}. Note that the smaller $\mathcal{M}_{ij}$ implies a better convergence of the channel square-norm to its long-term mean with high probability, and hence less fluctuation in the channel power around its long-term mean. Since the self-interference power is closely related to the channel power fluctuation, \emph{\textbf{decaying behavior of $\mathcal{M}_{ij}$ with the increasing AS implies reduction in the self-interference power, as well}}.

\section{Numerical Results and Discussion}\label{sec:numerical_results}
In this section, we present numerical results to evaluate the impact of AS on the achievable rates in a multi-cell network under pilot contamination and considering the EBF and the RZF precoders. The theoretical derivations presented in Section~\ref{sec:rates} are employed for analytical evaluations, and the corresponding simulation data is generated through extensive Monte Carlo runs. Without any loss of generality, we assume QPSK modulated pilot symbols in the UL training of sequence length $\tau\,{=}\,1$, the path-loss exponent $\gamma\,{=}\,3$, $\text{SNR}\,{=}\,0\text{ dB}$ with $\sigma^2\,{=}\,1$, $N_\mathrm{P}\,{=}\,100$, and $D\,{=}\,\frac{\lambda}{2}$ together with the distances $r_1\,{=}\,40\text{ m}$ and $r_2\,{=}\,50\text{ m}$ shown in Fig.~\ref{fig:layout}. Note here that rate results presented in this section are for the UE in $j$th cell.

\subsection{Two-Cell Scenario: Fixed Interfering UE Position}\label{sec:numerical_results_2cell_singlePos}

\begin{figure}
\begin{center}
\includegraphics[width=0.45\textwidth]{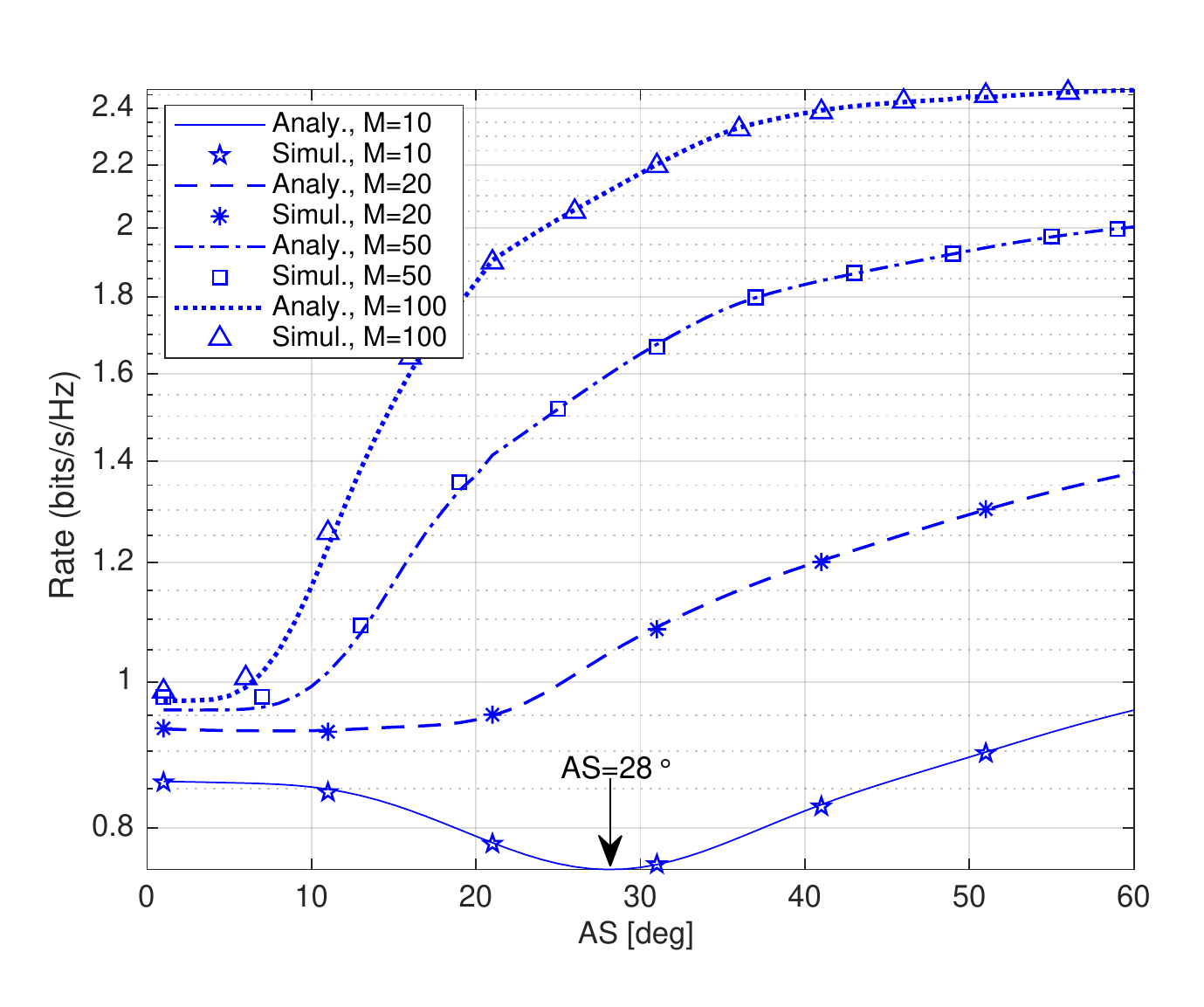}
\end{center}
\caption{Achievable rates for the EBF precoding in a $2$-cell scenario with $M\,{=}\,\{10,20,50,100\}$ and the interfering UE angular position at $\theta\,{=}\,200^{\circ}$.}
\label{fig:rate_Np=50_100_M=10_20}
\end{figure}

\begin{figure}
\begin{center}
\includegraphics[width=0.45\textwidth]{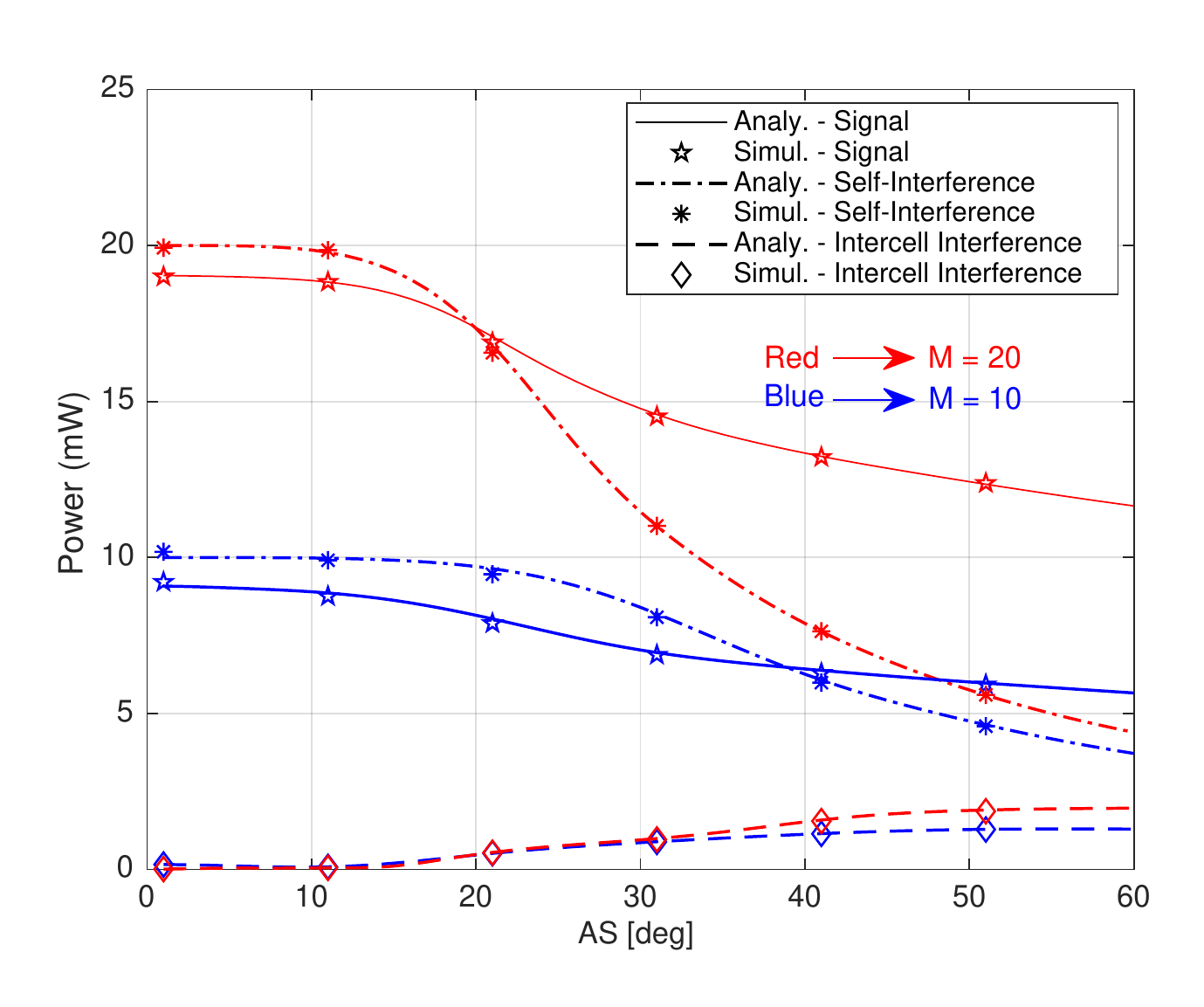}
\end{center}
\caption{Signal, self-interference and intercell interference powers for the EBF precoding in a $2$-cell scenario with $M\,{=}\,\{10,20\}$ and the interfering UE angular position at $\theta\,{=}\,200^{\circ}$.}
\label{fig:power_Np=50}
\end{figure}
This section considers a two-cell scenario where the $i$th and $j$th cells in Fig.~\ref{fig:layout} are designated as the interfering and the desired cells, respectively, and the angular position of the $i$th UE is $\theta\,{=}\,200^{\circ}$. Fig.~\ref{fig:rate_Np=50_100_M=10_20} captures the achievable rates with the EBF precoding for array sizes of $M\,{=}\,\{10,20,50,100\}$. We observe that the analytical results follow the characteristic behavior of the simulation data in all cases of interest. Further, we can observe from Fig.~\ref{fig:rate_Np=50_100_M=10_20} that, for $M\,{=}\,10$ rates are not monotonically increasing, and actually there is minimum rate value at $\text{AS}\,{=}\,28^{\circ}$. As it will become clear in Section~\ref{sec:numerical_results_multi_2cell_multiPos}, this \textit{unfavorable} AS value corresponding to the minimum rate depends on the underlying geometry. Therefore, the location of this minimum can be controlled through the deployment geometry, and, in particular through the angle of UE separation captured by $\Delta\theta$'s in Fig.~\ref{fig:layout}.

\smallskip

\begin{remark}
The real AS value of the propagation environment is independent of the underlying geometry, and it rather depends on the carrier frequency of the communication setting and some other features~\cite{Rappaport16mmWaveSta,3GPP16ChanMod}. As a result, the non-monotonic behavior of the achievable rates with respect to the AS (e.g. in Fig.~\ref{fig:rate_Np=50_100_M=10_20}) can be utilized to enhance the aggregate throughput. This can be achieved by discouraging the formation of user-cell pairs if the \textit{unfavorable} AS value associated with the minimum rate is close to the real AS value of the environment. To the best of our knowledge, none of the existing user-cell pairing approaches proposed in the literature exploit the AS of the propagation environment \cite{Yates_ULPwrControl_BS_Assignment, Rashid_DL_BS_assignment, Galeana_2008ABS, Galeana_CostBased_Approach_BS_Assignment,  Galeana_Backhaul_Aware_BS_Assign_OFDMA, 1687773}. Note that the \emph{non-monotonic behavior cannot be revealed through an asymptotic analysis} due to the moderately large antenna array size regime that we consider in this paper.
\end{remark}

\smallskip

\begin{remark}
Even though user rates for relatively larger antenna array sizes ($M=50$ and $M=100$) exhibit a sharp increase for small AS region, they tend to saturate eventually at larger AS values due to the severe pilot contamination, as discussed in Section~\ref{sec:channel_orthogonal}. On the other hand, relatively smaller 
array sizes ($M=10$ and $M=20$) result in no such saturation, which implies  that the pilot contamination is not a dominant effect over achievable rates in this array size regime.
\end{remark}

\smallskip

In Fig.~\ref{fig:power_Np=50}, the signal, the self-interference, and the intercell interference powers derived in Section~\ref{sec:rates} for the EBF precoder are captured separately for the scenario in Fig.~\ref{fig:rate_Np=50_100_M=10_20}. We observe that the analytical results follow the simulation data successfully in all cases of interest. The signal and the intercell interference  powers are observed to exhibit relatively flat characteristics over a range of small AS values up to approximately $10^{\circ}$. In this region, the AS values are sufficiently small, and the UL training beams of the UEs are therefore well separated. This is the reason for zero intercell interference in this region, which implies no pilot contamination effect and agrees with the geometrical interpretation of Section~\ref{sec:channel_orthogonal}.

When the AS increases beyond $10^{\circ}$, the intercell interference also starts increasing due to pilot contamination, and saturates around $\text{AS}\,{=}\,50^{\circ}$. Further, the DL transmission does not exactly align with the desired signal direction any more, which appears as the decreasing trend in the signal power as discussed in Section~\ref{sec:covariance_diagonal}. Note that, as captured in Fig.~\ref{fig:power_Np=50} the self-interference power has a decaying trend with the increasing AS as discussed in Section~\ref{sec:channel_fluctuation} and for $M\,{=}\,20$, self-interference starts decreasing earlier and much faster compared to $M\,{=}\,10$. Since the decrease in the signal power dominates initially for $M\,{=}\,10$ over self-interference, we observe a non-monotonic rate behavior for $M\,{=}\,10$ with a minimum at $\text{AS}\,{=}\,28^{\circ}$. On the other hand, since the self-interference starts decaying quickly for $M\,{=}\,20$ compared to the signal power, we observe monotonically increasing rate behavior for $M\,{\geq}\,20$.
\begin{figure}
\begin{center}
\includegraphics[width=0.45\textwidth]{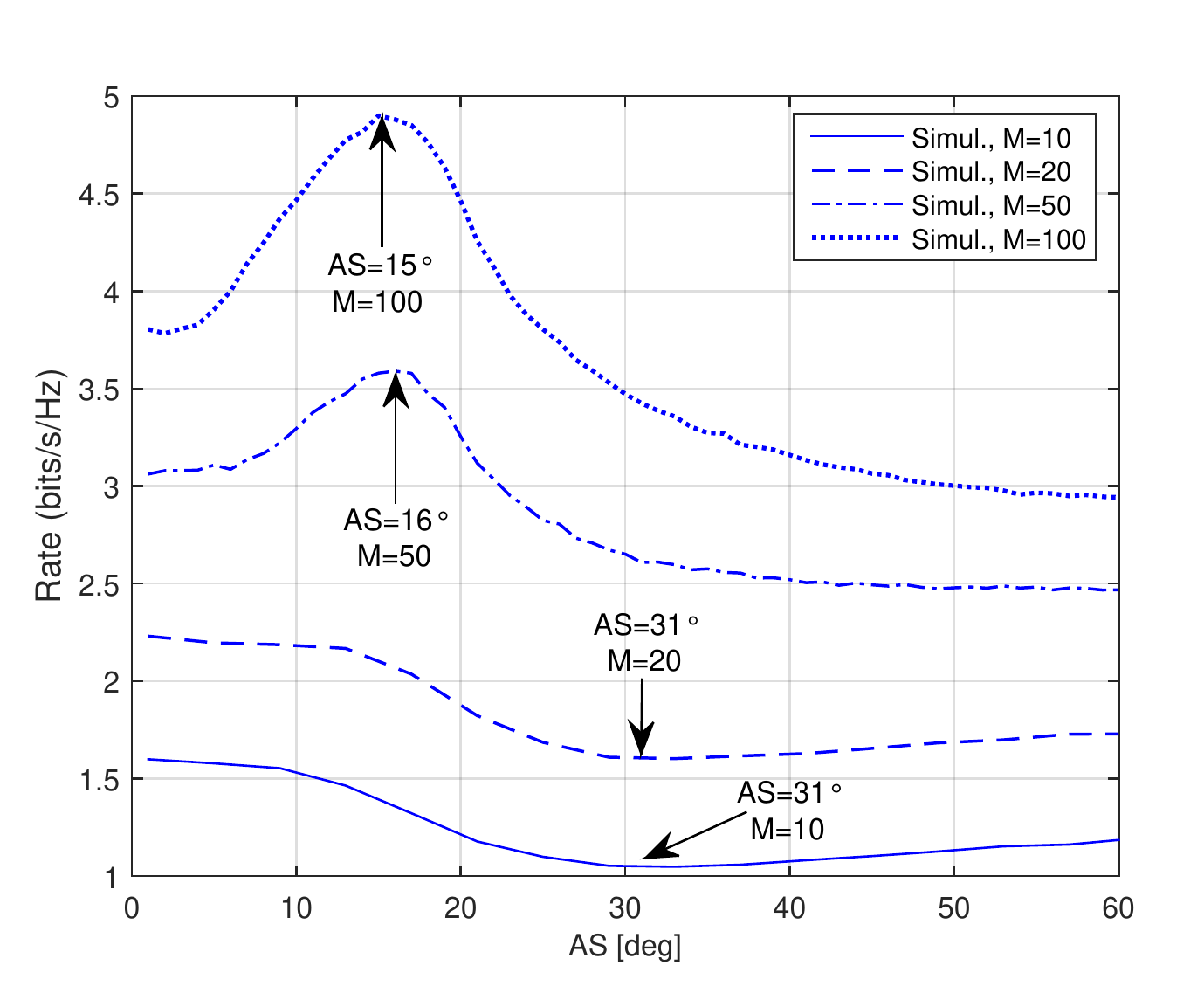}
\end{center}
\caption{Achievable rates for the RZF precoding in a $2$-cell scenario with $M\,{=}\,\{10,20,50,100\}$ and the interfering UE angular position at $\theta\,{=}\,200^{\circ}$.}
\label{fig:rate_rzf}
\end{figure}

The achievable rates for the RZF precoder in the DL transmission is depicted in Fig.~\ref{fig:rate_rzf} for the scenario of Fig.~\ref{fig:rate_Np=50_100_M=10_20} with the ULA sizes of $M\,{=}\,\{10,20,50,100\}$. We observe that the achievable rates with the RZF precoder is higher than that with the EBF precoder in Fig.~\ref{fig:rate_Np=50_100_M=10_20} for the same array sizes \emph{at the expense of a larger computational complexity}. We also observe a non-monotonic behavior in achievable rates for all the array sizes of interest, where there is a minimum at $\text{AS}\,{=}\,31^{\circ}$ for $M\,{=}\,\{10,20\}$, and a maximum at $\text{AS}\,{=}\,\{15^{\circ},16^{\circ}\}$ for $M\,{=}\,\{50,100\}$, respectively.

\smallskip

\begin{remark}
Similar to the EBF precoder case, the non-monotonic behavior as illustrated in Fig.~\ref{fig:rate_rzf} can be utilized effectively to enhance aggregate throughput by: 1) encouraging the formation of user-cell pairs if the favorable AS value associated with the maximum rate is close to the real AS value of the environment; and similarly, 2) discouraging the user-cell pairs for which the unfavorable AS value associated with the minimum rate is close to the real AS value of the environment.
\end{remark}

\smallskip

\subsection{Two-Cell Scenario: Varying Interfering UE Position}\label{sec:numerical_results_multi_2cell_multiPos}
\begin{figure}
\centering
\subfloat[EBF precoding]{\includegraphics[width=0.45\textwidth]{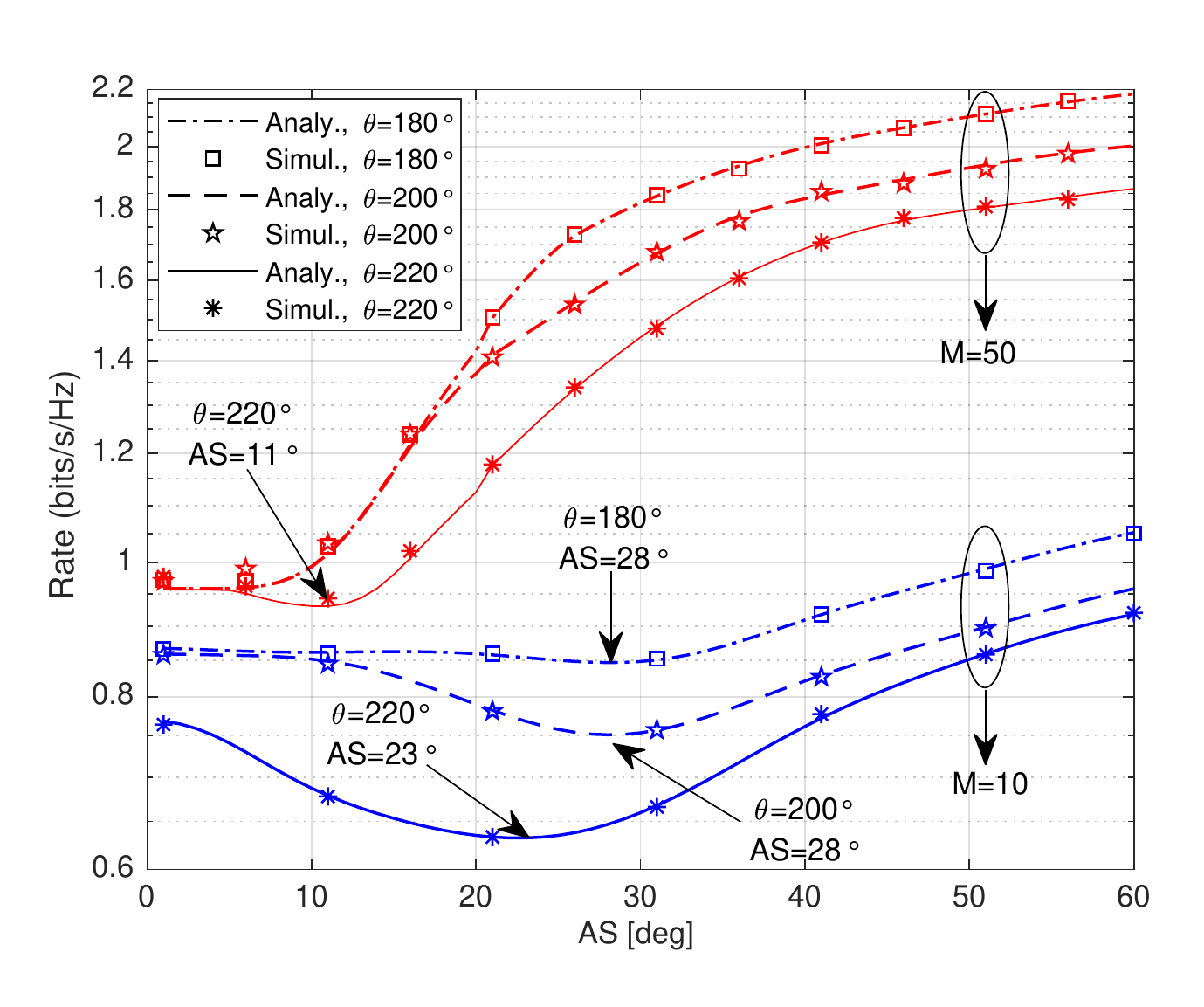}
\label{fig:rate_ebf_varyingTheta}}\\
\subfloat[RZF precoding]{\includegraphics[width=0.45\textwidth]{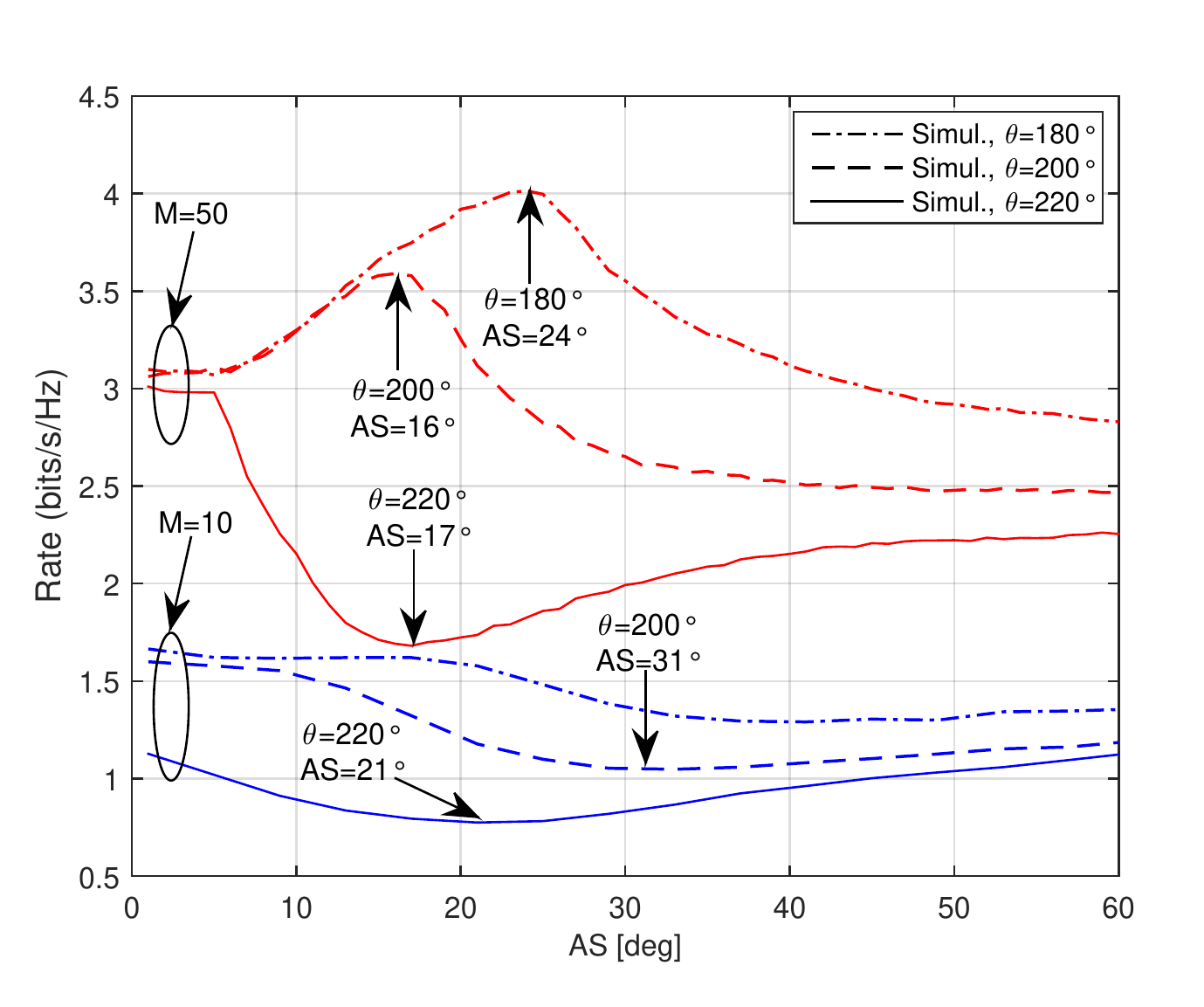}
\label{fig:rate_rzf_varyingTheta}}
\vspace{0.1in}
\caption{Achievable rates for the EBF and the RZF precoding in a $2$-cell scenario with $M\,{=}\,\{10,50\}$ and varying interfering UE position at $\theta\,{=}\,\{180^{\circ},200^{\circ},220^{\circ}\}$.}
\label{fig:rate_varyingTheta}
\end{figure}
\begin{figure}[!h]
\centering
\includegraphics[width=0.45\textwidth]{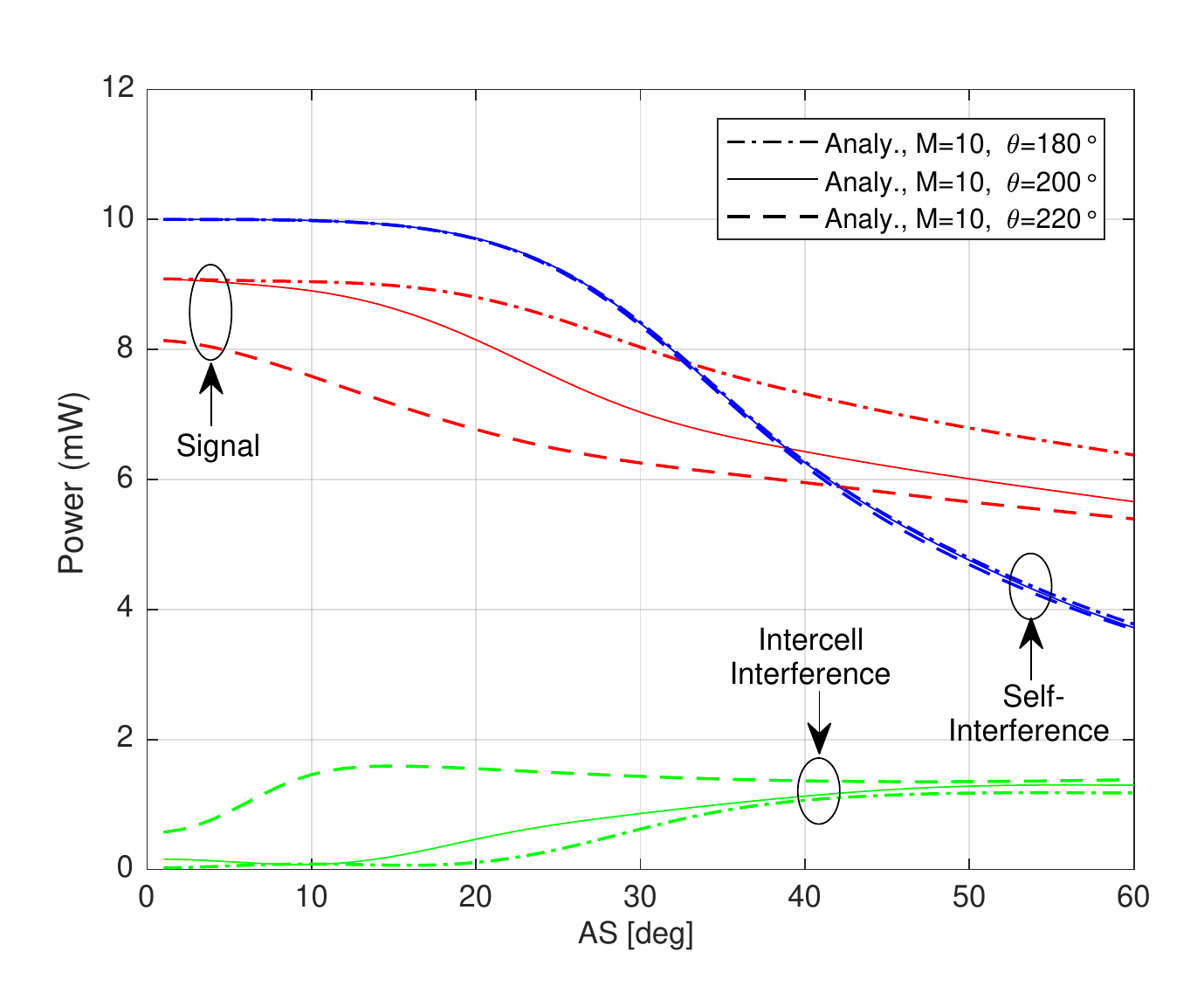}
\caption{Signal, self-interference and intercell interference powers for the EBF precoding in a $2$-cell scenario with $M\,{=}\,10$ and varying interfering UE positions at $\theta\,{=}\,\{180^{\circ},200^{\circ},220^{\circ}\}$.}
\label{fig:power_varyingTheta}
\end{figure}

In this section, we consider the effect of various angular positions of the $i$th interfering UE on achievable rates. Fig.~\ref{fig:rate_varyingTheta}, captures the achievable rates for the EBF and the RZF precoding in a $2$-cell scenario with $M\,{=}\,\{10,50\}$ at a set of angular positions $\theta\,{=}\,\{180^{\circ},200^{\circ},220^{\circ}\}$ for the $i$th interfering UE. We observe that as the interfering UE gets closer to the desired UE, which is indicated by the increasing $\theta$ in Fig.~\ref{fig:layout}, the achievable rate reduces for both the precoders. In addition, we observe either lower maxima or deeper minima located at smaller AS values, when $\theta$ increases.

As captured in Fig.~\ref{fig:power_varyingTheta}, the intercell interference increases for larger $\theta$ since the interfering UE gets closer to the desired UE and this indicates more powerful pilot contamination (see Section~\ref{sec:channel_orthogonal}).
In addition, the signal power gets smaller accordingly when $\theta$ increases since DL transmission does not align properly with the desired user direction any more, as discussed in Section~\ref{sec:covariance_diagonal}. Furthermore, the self-interference is not highly affected from $\theta$ (and hence from the pilot contamination) as shown in Fig.~\ref{fig:power_varyingTheta} along with the discussion in Section~\ref{sec:channel_fluctuation}. As a result, increasing intercell interference and decreasing signal power, both of which occur with increasing $\theta$, result in reduced user rates. This is also the reason behind the deeper minima observed for the EBF precoding for larger $\theta$ when $M\,{=}\,10$. Fig.~\ref{fig:rate_ebf_varyingTheta} shows that for $M\,{=}\,50$, the monotonically increasing rate behavior with the EBF precoding for $\theta\,{=}\,\{180^{\circ},200^{\circ}\}$ disappears for $\theta=220^{\circ}$, and instead a minimum value appears at $\text{AS}\,{=}\,11^\circ$. The intercell interference for larger $\theta$ can be very strong for $M\,{=}\,50$ with RZF such that the maxima at $\theta\,{=}\,\{180^{\circ},200^{\circ}\}$ switches into a minimum at $\theta\,{=}\,220^{\circ}$, as shown in Fig.~\ref{fig:rate_rzf_varyingTheta}.

\subsection{Multi-Cell Scenario}\label{sec:numerical_results_multi}

\begin{figure}
\centering
\subfloat[EBF precoding]{\includegraphics[width=0.45\textwidth]{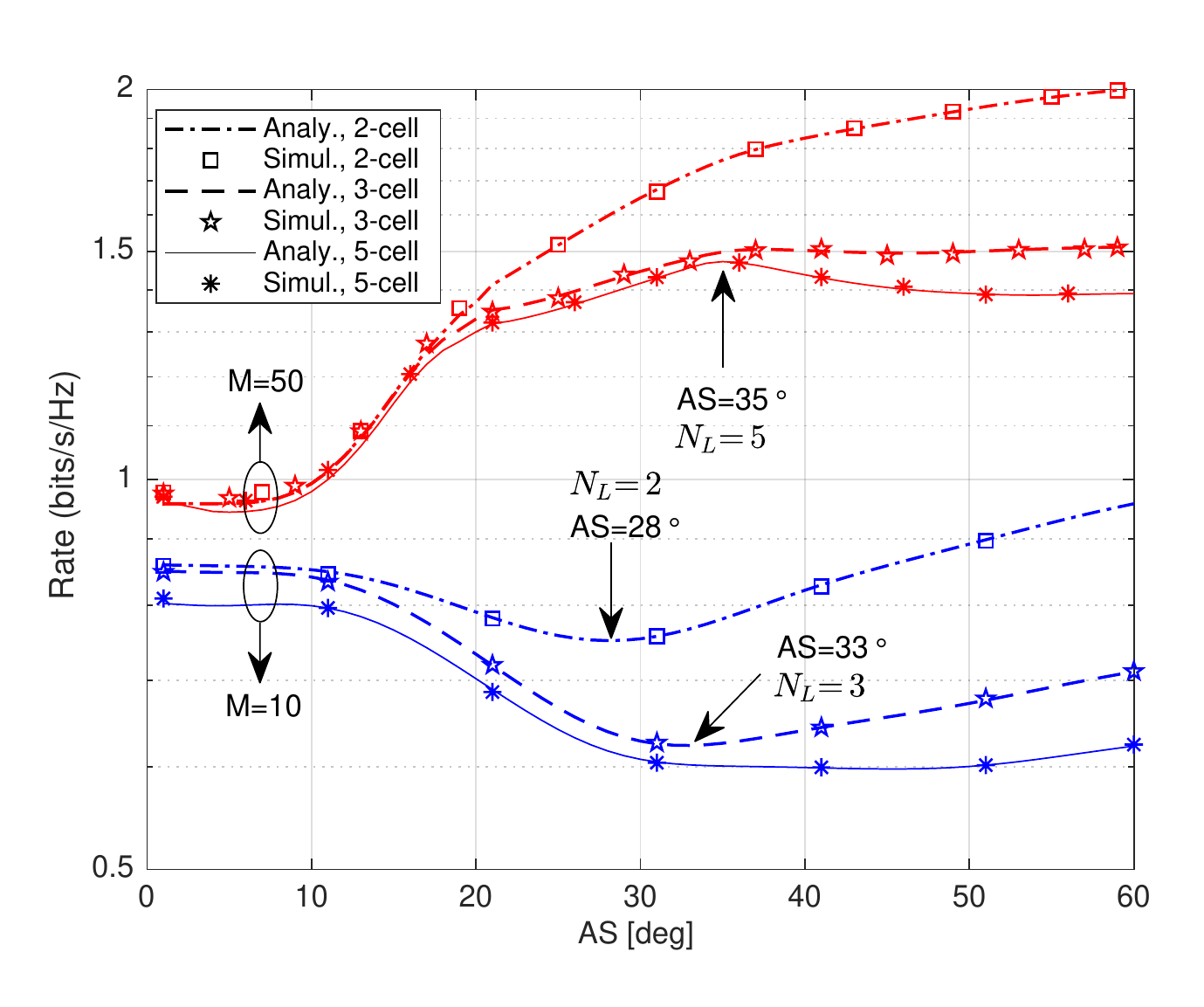}
\label{fig:rate_ebf_multipleCells}}\\
\subfloat[RZF precoding]{\includegraphics[width=0.45\textwidth]{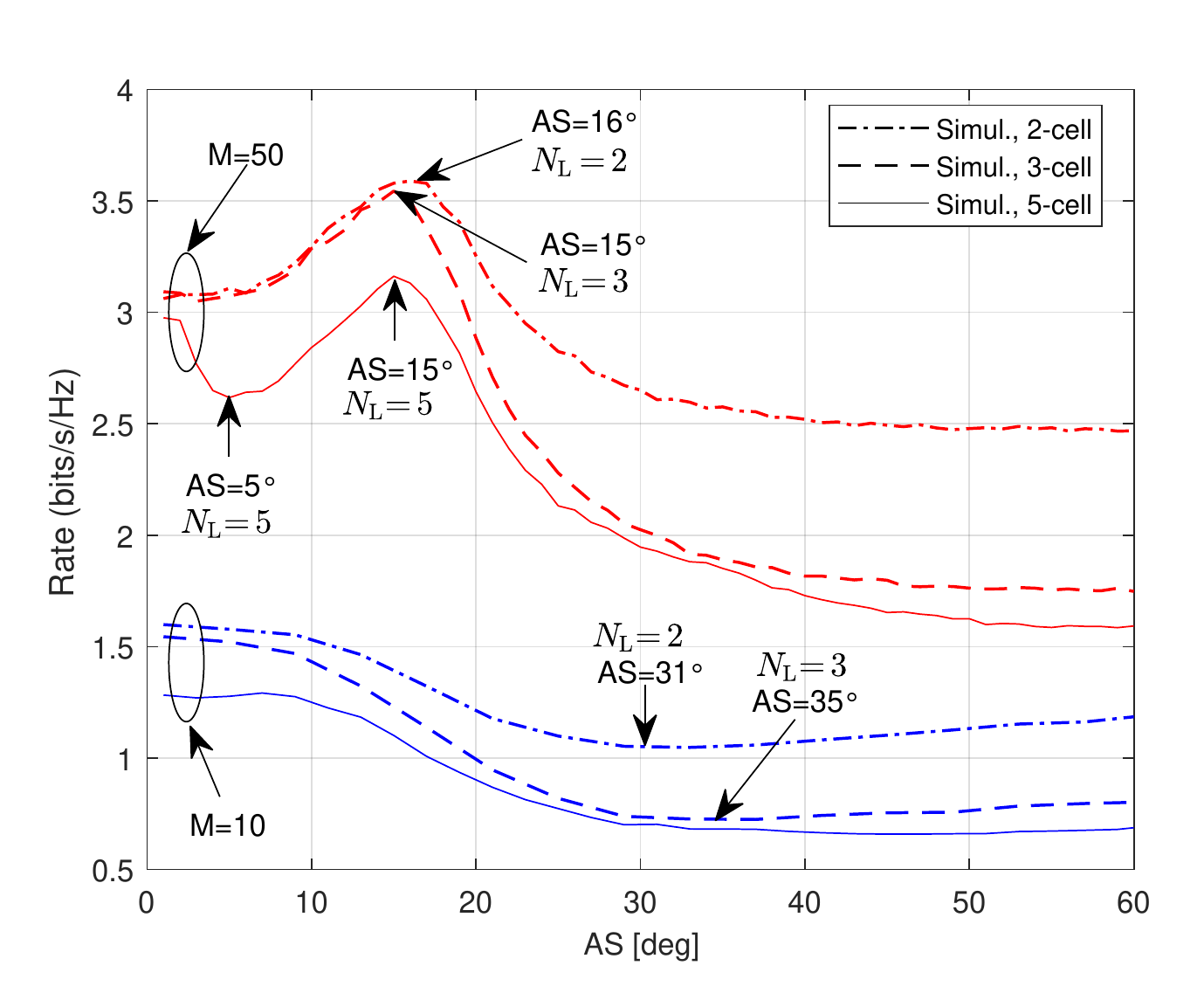}
\label{fig:rate_rzf_multipleCells}}
\vspace{0.1in}
\caption{Achievable rates for the EBF and the RZF precoding in a multi-cell scenario with $N_{\rm L}\,{=}\,\{2,3,5\}$, $M\,{=}\,\{10,50\}$.}
\label{fig:rate_multipleCells}
\end{figure}

\begin{figure}
\centering
\includegraphics[width=0.45\textwidth]{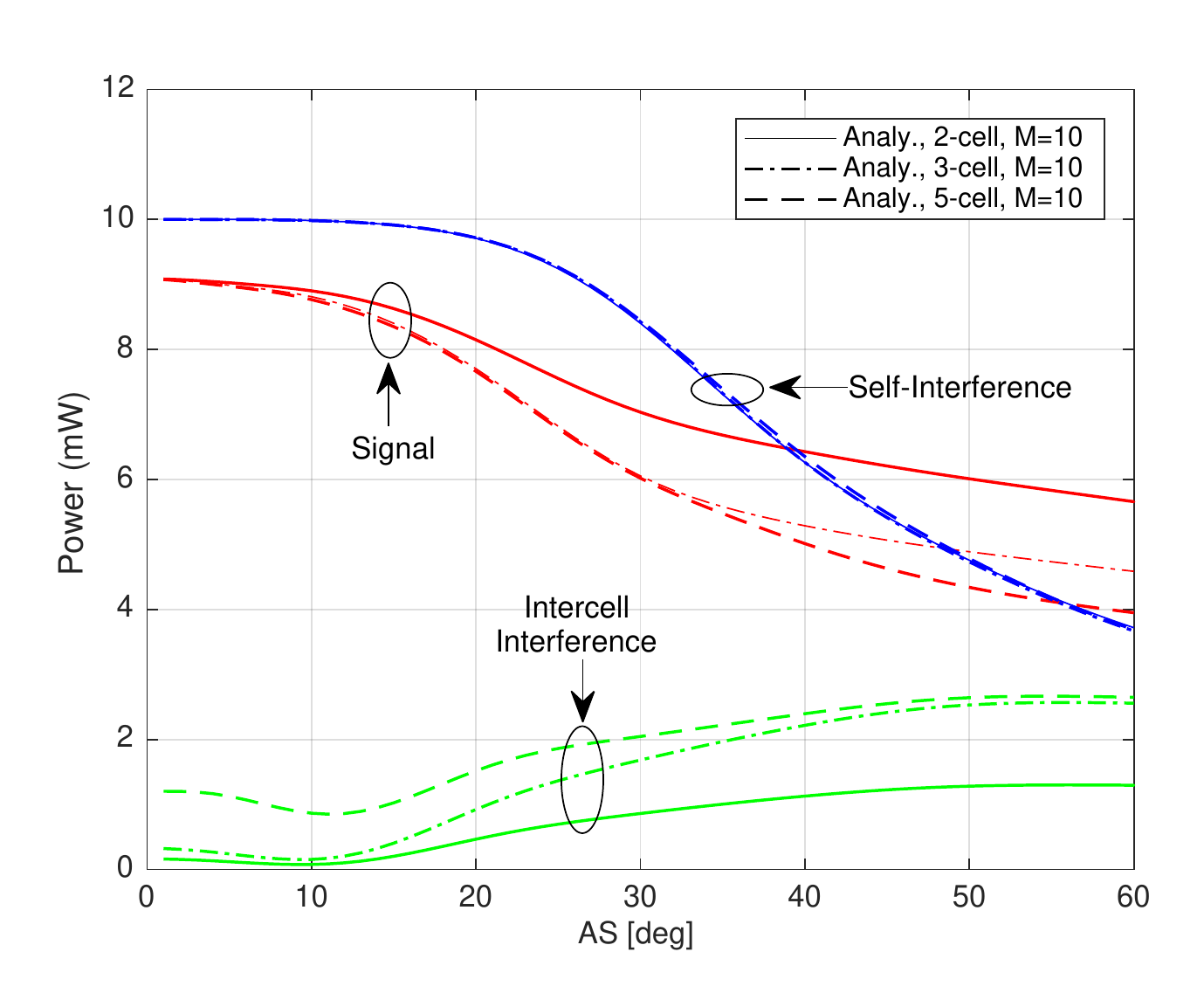}
\caption{Signal, self-interference and intercell interference powers for the EBF precoding in a multi-cell scenario with $N_{\rm L}\,{=}\,\{2,3,5\}$, $M\,{=}\,10$.}
\label{fig:power_multipleCells}
\end{figure}

Finally, we consider the impact of AS in a multi-cell setting with the number of cells $N_{\rm L}\,{\in}\,\{2,3,5\}$. To this end, a multi-cell setting is generated by considering five cells as shown in Fig.~\ref{fig:layout}, where the interfering UEs are located at $200^{\circ},160^{\circ},360^{\circ},60^{\circ}$ with respect to the horizontal axis for the $i$th, $k$th, $\ell$th, and $m$th cells, respectively. This layout provides almost the worst condition in terms of the intercell interference power, and hence the pilot contamination. The effect of this multi-cell setting on the achievable rates with the EBF and the RZF precoders is presented in Fig.~\ref{fig:rate_multipleCells} for $M\,{=}\,\{10,50\}$. We observe that as we consider more cells, the resulting interference  degrades achievable rates together with much lower maxima or deeper minima. We even observe the formation of an additional maximum for the EBF at $\text{AS}\,{=}\,35^\circ$ and minimum for the RZF at $\text{AS}\,{=}\,5^\circ$ when $N_{\rm L}\,{=}\,5$. Since adding more cells strengthens the intercell interference very rapidly, the desired signal power reduces proportionally, as shown in Fig.~\ref{fig:power_multipleCells}. These impairing effects eventually reduce and even saturate achievable rates, as shown in Fig.~\ref{fig:rate_multipleCells}.

\section{Concluding Remarks}\label{sec:conclusion}
We investigated the impact of AS on the achievable rates in a multi-cell environment under pilot contamination, considering moderately large antenna arrays. An exact analytical expression for achievable rate is derived for the EBF precoding considering arbitrary antenna array size. For correlated MIMO channels, we studied how interference channel orthogonality is affected from increasing AS along with the pilot contamination. Further, the channel power fluctuation around its long-term mean is analytically evaluated for varying ASs considering different antenna array sizes.

When the AS gets larger, we showed through rigorous analyses that 1) the covariance matrices tend to have a more diagonalized structure, 2) the channel power fluctuation diminishes (in a similar way as in the channel hardening), and 3) the orthogonality of the interference channel gets impaired due to pilot contamination effect. The overall achievable rate behavior as a function of the AS depends on which of these factors dominate over the other. Our analysis quantitatively identifies the antenna array size beyond which the pilot contamination starts being a dominant factor.

Lastly, our numerical results reveal a non-monotonic behavior (with respect to the AS) of the achievable rates 
for both the EBF and the RZF precoders under certain scenarios. The AS values at which the rate minimum/maximum occurs depend on the relative positions of the UEs and their serving BSs. Such a knowledge, along with the rate expression derived in this paper, can be effectively utilized to maximize the aggregate network throughput via careful design of user-cell pairing strategies. Due to space limitations we have left analytical rate evaluations with RZF precoder as a future research work.

\appendices
\section{Covariance Matrix Derivation}\label{app:cov_matrix}
The covariance matrix of the channel vector $\mathbf{h}_{ij}$ in \eqref{eqn:channel_defn} is given as
\small \begin{align}
\mathbf{R}_{ij} & = \frac{1}{N_\mathrm{P}} \sum\limits_{p=1}^{N_\mathrm{P}} \sum \limits_{p'=1}^{N_\mathrm{P}} \mathbb{E} \left\lbrace  \textbf{a}\left( \phi_{ij, p}\right)  \textbf{a}^{\rm H}\left( \phi_{ij, p'}\right) \right\rbrace \mathbb{E} \left\lbrace \alpha_{ij, p} \, \alpha_{ij, p'}^{*} \right\rbrace , \nonumber\\
& = \frac{\beta_{ij}}{N_\mathrm{P}} \sum \limits_{p=1}^{N_\mathrm{P}} \mathbb{E} \left\lbrace  \textbf{a}\left( \phi_{ij, p}\right)  \textbf{a}^{\rm H}\left( \phi_{ij, p}\right) \right\rbrace , \label{eqn:app:cov_matrix_1}\\
& = \beta_{ij} \mathbb{E} \left\lbrace  \textbf{a}\left( \phi_{ij}\right)  \textbf{a}^{\rm H}\left( \phi_{ij}\right) \right\rbrace~,  \label{eqn:app:cov_matrix_2}
\end{align} \normalsize
where \eqref{eqn:app:cov_matrix_1} employs $\mathbb{E} \left\lbrace \alpha_{ij, p}\,\alpha_{ij, p'}^{*} \right\rbrace\,{=}\,\beta_{ij} \delta(p,p')$, and \eqref{eqn:app:cov_matrix_2} follows from the fact that the distribution of AoA $\phi_{ij,p}$ is identical for any choice of the path index $p$. Defining the angular covariance matrix of the steering vector as $\mathbf{R}_{ij}^{\phi}\,{=}\,\mathbb{E}\left\lbrace \textbf{a}\left(\phi_{ij}\right) \textbf{a}^{\rm H}\left( \phi_{ij}\right) \right\rbrace$, and employing \eqref{eqn:steering_vector}, the element-wise angular correlation is given as follows
\begin{align}   \label{eqn:app:cov_phi_element}
& \mathbf{R}_{ij}^{\phi}(m,n) = \mathbb{E} \left\lbrace \exp \left( -j2\pi (m-n)\frac{D}{\lambda} \cos(\phi_{ij}) \right) \right\rbrace~, \nonumber \\
& = \int_{0}^{2\pi}  \exp \left( -j2\pi (m-n)\frac{D}{\lambda} \cos(\phi_{ij}) \right) p_{\phi}(\phi)\,{\rm d}\phi~,
\end{align} where $p_{\phi}(\phi)$ is the probability distribution function (pdf) of the AoA distribution. In particular, assuming the one-ring scatterer model \cite{Foschini_One_Ring_Scatterer,JSDM_LargeScaleArray} and uniform distribution for AoA with $\mathcal{U}\left[\bar{\phi}_{ij}{-}\Delta,\bar{\phi}_{ij}{+}\Delta\right]$, \eqref{eqn:app:cov_phi_element} can be given as

\small \begin{align}   \label{eqn:app:cov_phi_uni}
\mathbf{R}_{ij}^{\phi} (m,n) = \frac{1}{2\Delta}\int_{\bar{\phi}_{ij}-\Delta}^{\bar{\phi}_{ij}+\Delta}  \exp \left( -j2\pi (m-n)\frac{D}{\lambda} \cos(\phi_{ij}) \right) d\phi~.
\end{align}  \normalsize
Employing \eqref{eqn:app:cov_phi_element}, each entry of the covariance matrix in \eqref{eqn:app:cov_matrix_2} is given by \eqref{eqn:cov_matrix}. \hfill\IEEEQEDhere

We now derive the covariance matrix of the channel estimate $\hat{\textbf{h}}_{ij}$, denoted by $\hat{\textbf{R}}_{ij}\,{=}\, \mathbb{E} \left\lbrace {\hat{\textbf{h}}_{ij} \hat{\textbf{h}}_{ij}^{\rm H}} \right\rbrace $. Employing the definition of $\hat{\textbf{h}}_{ij}$ in \eqref{eqn:mmse_chan_est}, and the UL signal model in \eqref{eqn:UL_trans_vector}, $\hat{\textbf{R}}_{ij}$ is given as
\begin{align}
&\hat{\textbf{R}}_{ij} = \tilde{\textbf{R}}_{ij}\textbf{S}^{\rm H} \, \mathbb{E} \left\lbrace \textbf{y}_{j}^{\textrm{UL}} \left(\textbf{y}_{j}^{\textrm{UL}}\right)^{\rm H} \right\rbrace \textbf{S}\,\tilde{\textbf{R}}_{ij}^{\rm H} \,, \nonumber \\
&= \tilde{\textbf{R}}_{ij}\textbf{S}^{\rm H} \mathbb{E} \left\lbrace \left( \textbf{S} \sum \limits_{k=1}^{N_\mathrm{L}} \textbf{h}_{kj} +\textbf{n}_{j} \right) \left( \textbf{S} \sum \limits_{k=1}^{N_\mathrm{L}} \textbf{h}_{kj} +\textbf{n}_{j} \right)^{\rm H} \right\rbrace \textbf{S} \, \tilde{\textbf{R}}_{ij}^{\rm H}  \,,
\label{eqn:app:cov_est_1} \\
&= \tau^2 \sum \limits_{k=1}^{N_\mathrm{L}} \sum \limits_{\ell=1}^{N_\mathrm{L}} \tilde{\textbf{R}}_{ij}  \mathbb{E} \left\lbrace \textbf{h}_{kj}\textbf{h}_{\ell j}^{\rm H} \right\rbrace  \tilde{\textbf{R}}_{ij}^{\rm H}  +
\tau \sigma^2 \tilde{\textbf{R}}_{ij} \tilde{\textbf{R}}_{ij}^{\rm H}
\label{eqn:app:cov_est_2}  \\
& + \tau \sum \limits_{k=1}^{N_\mathrm{L}} \tilde{\textbf{R}}_{ij} \mathbb{E} \left\lbrace \textbf{h}_{kj} \textbf{n}_{j}^{\rm H} \right\rbrace \textbf{S} \tilde{\textbf{R}}_{ij}^{\rm H}
+ \tau \sum \limits_{k=1}^{N_\mathrm{L}} \tilde{\textbf{R}}_{ij}\textbf{S}^{\rm H} \mathbb{E} \left\lbrace \textbf{n}_{j} \textbf{h}_{kj}^{\rm H} \right\rbrace \tilde{\textbf{R}}_{ij}^{\rm H}  \,, \label{eqn:app:cov_est_3}
\end{align} \normalsize
where we employ the relations $\mathbb{E} \left\lbrace \textbf{n}_{j} \textbf{n}_{j}^{\rm H} \right\rbrace\,{=}\,\sigma^2\textbf{I}_{M\tau}$ and $\textbf{S}^H\textbf{S}\,{=}\,\tau \textbf{I}_M$. Since the noise and the channel vectors are uncorrelated and zero-mean, the expectations in \eqref{eqn:app:cov_est_3} cancel, and we have
\begin{align}
\hat{\textbf{R}}_{ij} &= \tau^2 \sum \limits_{k=1}^{N_\mathrm{L}} \tilde{\textbf{R}}_{ij} \mathbb{E}\left\lbrace \textbf{h}_{kj}\textbf{h}_{kj}^{\rm H} \right\rbrace \tilde{\textbf{R}}_{ij}^{\rm H}  \nonumber \\
&+ \tau^2 \sum\limits_{k=1}^{N_\mathrm{L}} \sum \limits_{\substack{\ell=1 \\ \ell \neq k}}^{N_\mathrm{L}} \tilde{\textbf{R}}_{ij} \mathbb{E}\left\lbrace \textbf{h}_{kj}\textbf{h}_{\ell j}^{\rm H} \right\rbrace \tilde{\textbf{R}}_{ij}^{\rm H} + \tau \sigma^2 \tilde{\textbf{R}}_{ij} \tilde{\textbf{R}}_{ij}^{\rm H} \,, \label{eqn:app:cov_est_4}
\end{align}
where the second term in \eqref{eqn:app:cov_est_4} vanishes since $\mathbb{E}\left\lbrace\textbf{h}_{kj}\textbf{h}_{\ell j}^{\rm H}\right\rbrace\,{=}\,\textbf{0}_M$ for $k\,{\neq}\,\ell$. Then $\hat{\textbf{R}}_{ij} $ becomes
\begin{align}
\hat{\textbf{R}}_{ij} &= \sum \limits_{k=1}^{N_\mathrm{L}} \tau^2 \tilde{\textbf{R}}_{ij} \textbf{R}_{kj} \tilde{\textbf{R}}_{ij}^{\rm H} + \tau \sigma^2 \tilde{\textbf{R}}_{ij} \tilde{\textbf{R}}_{ij}^{\rm H} \,, \nonumber \\
&= \tau \tilde{\textbf{R}}_{ij} \left( \tau \sum \limits_{k=1}^{N_\mathrm{L}}  \textbf{R}_{kj} + \sigma^2 \textbf{I}_{M} \right) \tilde{\textbf{R}}_{ij}^{\rm H} = \tau \textbf{R}_{ij} \tilde{\textbf{R}}_{ij}^{\rm H} \,, \label{eqn:app:cov_est_6}
\end{align} \normalsize
where we employ hermitian symmetry of covariance matrices $\textbf{R}_{ij} $ and $\hat{\textbf{R}}_{ij}$ to obtain \eqref{eqn:cov_matrix_mmse}.  \hfill\IEEEQEDhere

\section{First and Second Order Moment Derivation}\label{app:achievable_rates} 
In this section, we derive the first and second order moments $\mathbb{E}\left\lbrace\textbf{h}_{jj}^{\rm H}\textbf{w}_j  \right\rbrace$ and $\mathbb{E} \left\lbrace \left| \textbf{h}_{ji}^{\rm H} \textbf{w}_i \right|^2 \right\rbrace$, respectively, for the EBF precoding given in \eqref{eqn:ebf}. Before the analysis, we define the following property which is used throughout this section while evaluating the mean of the quadratic and the double-quadratic forms involving random vectors.

\begin{lemma}\label{lem:quadratic}
Assume that $\left\lbrace \textbf{u}_i \right\rbrace_{i=1}^{4}$ be a set of zero-mean random vectors of arbitrary sizes where each of them may be individually correlated with the arbitrary covariance matrices $\left\lbrace \textbf{C}_i \right\rbrace_{i=1}^{4}$. For the given coefficient matrices $\textbf{A}$ and $\textbf{B}$ of the appropriate sizes and with arbitrary entries, the quadratic form $\textbf{u}_1^{\rm H} \textbf{A} \textbf{u}_2$ and the double-quadratic form $\textbf{u}_1^{\rm H} \textbf{A} \textbf{u}_2 \textbf{u}_3^{\rm H} \textbf{B} \textbf{u}_4$ are zero-mean if at least one of these random vectors are uncorrelated with the others.
\end{lemma}

\begin{IEEEproof}
Assuming that $\textbf{u}_1$ is uncorrelated with the others, without any loss of generality, regardless of whether $\left\lbrace \textbf{u}_i \right\rbrace_{i=2}^{4}$ are correlated with each other or not, we have
\begin{align*}
&\mathbb{E}\left\lbrace \textbf{u}_1^{\rm H} \textbf{A} \textbf{u}_2 \right\rbrace = \sum_{m}\sum_{n} \textbf{A}(m,n) \mathbb{E}\left\lbrace u_{1,m}^{\rm *} \right\rbrace \mathbb{E}\left\lbrace u_{2,n} \right\rbrace = 0~,\\
&\mathbb{E}\left\lbrace \textbf{u}_1^{\rm H} \textbf{A} \textbf{u}_2 \textbf{u}_3^{\rm H} \textbf{B} \textbf{u}_4 \right\rbrace = \sum_{m}\sum_{n}\sum_{k}\sum_{\ell} \textbf{A}(m,n)\textbf{B}(k,\ell) \times \\
& \hspace{10em} \mathbb{E}\left\lbrace u_{1,m}^{\rm *} \right\rbrace \mathbb{E}\left\lbrace u_{2,n} u_{3,k}^{\rm *} u_{4,\ell} \right\rbrace = 0~,
\end{align*} where $u_{i,m}$ denotes the $m$th entry of $\textbf{u}_{i}$.
\end{IEEEproof}
\subsection{First Order Moment}\label{app:1st_moment}
Employing the UL signal model in \eqref{eqn:UL_trans_vector} and the channel estimate in \eqref{eqn:mmse_chan_est}, the first order moment of the desired signal is given as follows
\begin{align}
&\mathbb{E} \left\lbrace \textbf{h}_{jj}^{\rm H} \textbf{w}_j  \right\rbrace = \mathbb{E} \left\lbrace \textbf{h}_{jj}^{\rm H} \tilde{\textbf{R}}_{jj}\textbf{S}^{\rm H} \left( \textbf{S} \sum \limits_{i=1}^{N_\mathrm{L}} \textbf{h}_{ij} +\textbf{n}_{j} \right) \right\rbrace \,,
\nonumber \\
&= \tau \sum \limits_{i=1}^{N_\mathrm{L}} \mathbb{E} \left\lbrace \textbf{h}_{jj}^{\rm H} \tilde{\textbf{R}}_{jj} \textbf{h}_{ij} \right\rbrace + \mathbb{E} \left\lbrace \textbf{h}_{jj}^{\rm H} \tilde{\textbf{R}}_{jj} \textbf{S}^{\rm H} \textbf{n}_{j}  \right\rbrace \,,\label{eqn:app:1st_moment_1}
\end{align}
where the last line employs $\textbf{S}^H\textbf{S}\,{=}\,\tau \textbf{I}_M$. Since $\textbf{h}_{jj}$ and $\textbf{n}_{j}$ are zero-mean and uncorrelated, the second expectation in \eqref{eqn:app:1st_moment_1} vanishes as per Lemma~\ref{lem:quadratic}. The desired expectation becomes
\begin{align}
\mathbb{E} \left\lbrace \textbf{h}_{jj}^{\rm H} \textbf{w}_j  \right\rbrace &= \tau \mathbb{E} \left\lbrace \textbf{h}_{jj}^{\rm H} \tilde{\textbf{R}}_{jj}  \textbf{h}_{jj} \right\rbrace {+} \tau \sum \limits_{\substack{i=1\\i \neq j}}^{N_\mathrm{L}} \mathbb{E} \left\lbrace \textbf{h}_{jj}^{\rm H} \tilde{\textbf{R}}_{jj}  \textbf{h}_{ij} \right\rbrace ,\label{eqn:app:1st_moment_2}\\
&= \tau \mathbb{E} \left\lbrace \textbf{h}_{jj}^{\rm H} \tilde{\textbf{R}}_{jj}  \textbf{h}_{jj} \right\rbrace\,, \label{eqn:app:1st_moment_3}
\end{align}
where the second expectation in \eqref{eqn:app:1st_moment_2} is similarly zero as per Lemma~\ref{lem:quadratic} since the channels of $i$th and $j$th UEs to the $j$th BS, denoted by $\textbf{h}_{ij}$ and $\textbf{h}_{jj}$, respectively, are uncorrelated from each other, and zero-mean by definition. Finally, representing \eqref{eqn:app:1st_moment_3} by using the trace operator, employing the Hermitian symmetry of the covariance matrix, and incorporating the covariance matrix of the channel estimate in \eqref{eqn:cov_matrix_mmse} yield the desired expression given in \eqref{eqn:1st_moment}.
\subsection{Second Order Moment}\label{app:2nd_moment}
Employing \eqref{eqn:UL_trans_vector} and \eqref{eqn:mmse_chan_est}, as in Appendix~\ref{app:1st_moment}, the second order moment of the desired signal is given as follows

 \small \begin{align}
&\mathbb{E} \left\lbrace \left| \textbf{h}_{ji}^{\rm H} \textbf{w}_i \right|^2 \right\rbrace = \mathbb{E} \left\lbrace \textbf{h}_{ji}^{\rm H} \hat{\textbf{h}}_{ii} \hat{\textbf{h}}_{ii}^{\rm H} \textbf{h}_{ji} \right\rbrace \,, \\
&= \mathbb{E} \Bigg\{ \textbf{h}_{ji}^{\rm H} \tilde{\textbf{R}}_{ii}\textbf{S}^{\rm H} \left( \textbf{S} \sum \limits_{k=1}^{N_\mathrm{L}} \textbf{h}_{ki} +\textbf{n}_{i} \right) \left( \textbf{S} \sum \limits_{\ell=1}^{N_\mathrm{L}} \textbf{h}_{\ell i} +\textbf{n}_{i} \right)^{\rm H} \textbf{S} \tilde{\textbf{R}}_{ii}^{\rm H} \textbf{h}_{ji} \Bigg\} \,, \\
&=  \tau^2  \sum \limits_{k=1}^{N_\mathrm{L}} \sum \limits_{\ell=1}^{N_\mathrm{L}} \mathbb{E} \left\lbrace \textbf{h}_{ji}^{\rm H} \tilde{\textbf{R}}_{ii} \textbf{h}_{ki} \textbf{h}_{\ell i}^{\rm H} \tilde{\textbf{R}}_{ii}^{\rm H} \textbf{h}_{ji} \right\rbrace +  \mathbb{E} \left\lbrace \textbf{h}_{ji}^{\rm H} \tilde{\textbf{R}}_{ii} \textbf{S}^{\rm H} \textbf{n}_{i} \textbf{n}_{i}^{\rm H} \textbf{S} \tilde{\textbf{R}}_{ii}^{\rm H} \textbf{h}_{ji} \right\rbrace \\
&+ \tau  \sum \limits_{k=1}^{N_\mathrm{L}} \mathbb{E} \left\lbrace \textbf{h}_{ji}^{\rm H} \tilde{\textbf{R}}_{ii} \textbf{S}^{\rm H} \textbf{n}_{i} \textbf{h}_{ki}^{\rm H} \tilde{\textbf{R}}_{ii}^{\rm H} \textbf{h}_{ji} \right\rbrace  +  \tau \sum \limits_{k=1}^{N_\mathrm{L}}  \mathbb{E} \left\lbrace \textbf{h}_{ji}^{\rm H} \tilde{\textbf{R}}_{ii} \textbf{h}_{ki} \textbf{n}_{i}^{\rm H} \textbf{S} \tilde{\textbf{R}}_{ii}^{\rm H} \textbf{h}_{ji} \right\rbrace \,, \label{eqn:app:2nd_moment_1}
\end{align} \normalsize
where the expectations in \eqref{eqn:app:2nd_moment_1} vanishes in accordance with Lemma~\ref{lem:quadratic} since $\textbf{n}_{i}$ is uncorrelated with $\textbf{h}_{ji}$ and $\textbf{h}_{ki}$, and hence
\begin{align}\label{eqn:app:2nd_moment_2}
\mathbb{E} \left\lbrace \left| \textbf{h}_{ji}^{\rm H} \textbf{w}_i \right|^2 \right\rbrace &= \underbrace{ \tau^2  \sum \limits_{k=1}^{N_\mathrm{L}} \sum\limits_{\ell=1}^{N_\mathrm{L}} \mathbb{E} \left\lbrace \textbf{h}_{ji}^{\rm H} \tilde{\textbf{R}}_{ii} \textbf{h}_{ki} \textbf{h}_{\ell i}^{\rm H} \tilde{\textbf{R}}_{ii}^{\rm H} \textbf{h}_{ji} \right\rbrace}_{\textmd{${\rm E}_1$}}
\\ \nonumber
& \hspace{15 mm} +  \underbrace{ \vphantom{ \sum \limits_{k=1}^{N_\mathrm{L}} } \mathbb{E} \left\lbrace \textbf{h}_{ji}^{\rm H} \tilde{\textbf{R}}_{ii} \textbf{S}^{\rm H} \textbf{n}_{i} \textbf{n}_{i}^{\rm H} \textbf{S} \tilde{\textbf{R}}_{ii}^{\rm H} \textbf{h}_{ji} \right\rbrace }_{\textmd{${\rm E}_2$}}~.
\end{align}
In the following, we will elaborate the two expectations, ${\rm E}_1$ and ${\rm E}_2$, in \eqref{eqn:app:2nd_moment_2}, separately. We start with ${\rm E}_1$ as follows
\begin{align}\label{eqn:app:2nd_moment_e1_1}
{\rm E}_1 &= \tau^2  \sum \limits_{k=1}^{N_\mathrm{L}}  \mathbb{E} \left\lbrace \textbf{h}_{ji}^{\rm H} \tilde{\textbf{R}}_{ii} \textbf{h}_{ki} \textbf{h}_{ki}^{\rm H} \tilde{\textbf{R}}_{ii}^{\rm H} \textbf{h}_{ji} \right\rbrace
\\ \nonumber
& \hspace{20 mm} + \tau^2  \sum \limits_{k=1}^{N_\mathrm{L}} \sum \limits_{\substack{\ell=1\\ \ell \neq k}}^{N_\mathrm{L}} \mathbb{E} \left\lbrace \textbf{h}_{ji}^{\rm H} \tilde{\textbf{R}}_{ii} \textbf{h}_{ki} \textbf{h}_{\ell i}^{\rm H} \tilde{\textbf{R}}_{ii}^{\rm H} \textbf{h}_{ji} \right\rbrace~,
\end{align}
where $\textbf{h}_{ki}$ and $\textbf{h}_{\ell i}$ in the second expectation are obviously uncorrelated as $k \neq \ell$. Note that, $\textbf{h}_{ki}$ is uncorrelated with $\textbf{h}_{\ell i}$ and $\textbf{h}_{ji}$ when $j=\ell$, and $\textbf{h}_{\ell i}$ is uncorrelated with $\textbf{h}_{ki}$ and $\textbf{h}_{ji}$ when $j=k$, and finally all $\textbf{h}_{ki}$, $\textbf{h}_{\ell i}$ and $\textbf{h}_{ji}$ are uncorrelated when $j\neq\{k,\ell\}$. As a result, in any case, we have at least one zero-mean vector uncorrelated with the others, and the second expectation in \eqref{eqn:app:2nd_moment_e1_1} is therefore zero in accordance with Lemma~\ref{lem:quadratic}. As a result, ${\rm E}_1$ in \eqref{eqn:app:2nd_moment_e1_1} becomes

\vspace{-2.1em}
\small \begin{align}
&{\rm E}_1= \tau^2  \sum \limits_{k=1}^{N_\mathrm{L}}  \mathbb{E} \left\lbrace \textbf{h}_{ji}^{\rm H} \tilde{\textbf{R}}_{ii} \textbf{h}_{ki} \textbf{h}_{ki}^{\rm H} \tilde{\textbf{R}}_{ii}^{\rm H} \textbf{h}_{ji} \right\rbrace \nonumber \,,\\
&= \tau^2 \underbrace{\mathbb{E} \left\lbrace \textbf{h}_{ji}^{\rm H} \tilde{\textbf{R}}_{ii} \textbf{h}_{ji} \textbf{h}_{ji}^{\rm H} \tilde{\textbf{R}}_{ii}^{\rm H} \textbf{h}_{ji} \right\rbrace}_{{\rm E}_{11}} + \tau^2 \sum \limits_{\substack{k=1\\ k \neq j}}^{N_\mathrm{L}} \underbrace{ \mathbb{E} \left\lbrace \textbf{h}_{ji}^{\rm H} \tilde{\textbf{R}}_{ii} \textbf{h}_{ki} \textbf{h}_{ki}^{\rm H} \tilde{\textbf{R}}_{ii}^{\rm H} \textbf{h}_{ji} \right\rbrace}_{{\rm E}_{12}}\,, \label{eqn:app:2nd_moment_e1_2}
\end{align} \normalsize
and the first expectation ${\rm E}_{11}$ can be expressed in weighted sum of scalars as follows
\begin{align} \label{eqn:app:2nd_moment_e1_3}
{\rm E}_{11}\, &{=}\,\sum\limits_{m=1}^{M} \sum\limits_{n=1}^{M} \sum\limits_{{m'}=1}^{M} \sum\limits_{{n'}=1}^{M} \tilde{\textbf{R}}_{ii}(m,n) \, \tilde{\textbf{R}}_{ii}^{\rm H}({m'},{n'}) \, \times
\\ \nonumber
& \hspace{10em} \underbrace{\mathbb{E} \left\lbrace h_{ji,m}^{\ast}h_{ji,n} h_{ji,{m'}}^{\ast} h_{ji,{n'}} \right\rbrace}_{{\rm E}_{\phi}(m,n,{m'},{n'})}~,
\end{align} where $h_{ji,m}$ is the $m$th element of the channel vector $\textbf{h}_{ji}$, and is given by employing \eqref{eqn:channel_defn} and \eqref{eqn:steering_vector} as follows

\small \begin{align} \label{eqn:channel_m}
h_{ji,m} = \dfrac{1}{\sqrt{N_\mathrm{P}}}\sum \limits_{p=1}^{N_\mathrm{P}} \alpha_{ji,p} \exp\left\lbrace {-}j 2\pi \frac{D}{\lambda}(m-1)\cos\left(\phi_{ji,p}\right) \right\rbrace~.
\end{align} \normalsize By \eqref{eqn:channel_m}, the expectation at the right-hand side of \eqref{eqn:app:2nd_moment_e1_3} can be further elaborated as follows
\begin{align}\label{eqn:app:2nd_moment_e1_4}
{\rm E}_{\phi} &(m,n,{m'},{n'})\,{=}\,\dfrac{1}{{N_\mathrm{P}}^2} \sum \limits_{p_1{=}1}^{N_\mathrm{P}} \sum \limits_{p_2{=}1}^{N_\mathrm{P}} \sum \limits_{p_3{=}1}^{N_\mathrm{P}} \sum \limits_{p_4{=}1}^{N_\mathrm{P}} \mathbb{E}_{\alpha} \times
\\ \nonumber
& \hspace{5 mm}  \mathbb{E} \left\lbrace \exp \left( {-}j 2\pi \frac{D}{\lambda} \sum\limits_{\nu{=}1}^{4}({-}1)^{\nu}\left(u_{\nu}{-}1\right)\cos\left(\phi_{ji,p_{\nu}}\right) \right) \right\rbrace ,
\end{align}
$\left\lbrace u_{\nu}\right\rbrace_{\nu{=}1}^{4}{=}\{m,n,{m'},{n'}\}$, $\mathbb{E}_{\alpha}\,{=}\,\mathbb{E} \left\lbrace  \alpha_{ji, p_1}^{\ast} \alpha_{ji, p_2} \alpha_{ji, p_3}^{\ast} \alpha_{ji, p_4} \right\rbrace$. Note that, $\mathbb{E}_{\alpha}$ is nonzero only when 1) $p_1\,{=}\,p_2\,{=}\,p_3\,{=}\,p_4$, 2) $p_1\,{=}\,p_2, p_3\,{=}\,p_4$ (with $p_1\,{\neq}\,p_3)$, or 3) $p_1\,{=}\,p_4, p_2\,{=}\,p_3$ (with $p_1\,{\neq}\,p_2)$, and zero otherwise, since $\alpha_{ji,p}$ is zero-mean and uncorrelated over the path index $p$. Next, we analyze these three conditions to have a closed-form expression for \eqref{eqn:app:2nd_moment_e1_4}.

\begin{remark}
Note that, the other possibilities for the path indices $\left\lbrace p_{\nu} \right\rbrace_{\nu{=}1}^{4}$ for which $\mathbb{E}_{\alpha}$ is zero, consist of the cases where i) none of the path indices equal to the other, ii) one of the path indices is not equal to all the others, and iii) the pairwise equality with $p_1\,{=}\,p_3, p_2\,{=}\,p_4$. For the cases i) and ii), the expectation $\mathbb{E}_{\alpha}$ involves a term $\left[\mathbb{E}\left\lbrace \alpha_{ji, p} \right\rbrace\right]^\kappa$ with $\kappa\,{\geq}\,1$ which is zero since $\alpha_{ji, p}$ is zero-mean, and hence yields $\mathbb{E}_{\alpha}\,{=}\,0$. The case iii) yields $\mathbb{E}_{\alpha}\,{=}\,\left|\mathbb{E} \left\lbrace \alpha_{ji,p}^2\right\rbrace \right|^2$ which can easily be shown to be zero as $\alpha_{ji, p}$ has uncorrelated real and imaginary parts which are zero-mean.
\end{remark}

\begin{case}\label{case:case_1}
Assuming $p_1\,{=}\,p_2\,{=}\,p_3\,{=}\,p_4$, the desired expectation ${\rm E}_{\phi}(m,n,{m'},{n'})$ in \eqref{eqn:app:2nd_moment_e1_4} becomes
\begin{align}
&{\rm E}_{\phi}(m,n,{m'},{n'}) = \dfrac{1}{{N_\mathrm{P}}^2} \sum\limits_{p=1}^{N_\mathrm{P}} \mathbb{E} \left\lbrace \left| \alpha_{ji,p} \right|^4 \right\rbrace \times
\nonumber \\
& \hspace{4em} \mathbb{E} \left\lbrace \exp\left( {-}j 2\pi \frac{D}{\lambda} \left( n{-}m{+}{n'}{-}{m'} \right) \cos\left(\phi_{ji,p}\right) \right) \right\rbrace \,,\nonumber \\
&\stackrel{\text{(a)}}{=} \dfrac{1}{{N_\mathrm{P}}} \mathbb{E} \left\lbrace \left| \alpha_{ji} \right|^4 \right\rbrace \times
\nonumber \\
& \hspace{4em} \mathbb{E} \left\lbrace \exp\left( {-}j 2\pi \frac{D}{\lambda} \left( n{-}m{+}{n'}{-}{m'} \right) \cos\left(\phi_{ji}\right) \right) \right\rbrace
\,,\nonumber \\
&\stackrel{\text{(b)}}{=}\dfrac{ 2\beta_{ji}^2 }{{N_\mathrm{P}}} {\rm E}_{ji}( n{-}m{+}{n'}{-}{m'} )\,, \label{eqn:app:2nd_moment_e1_5}
\end{align}
where $\text{(a)}$ follows from the uncorrelatedness of $\alpha _{ji}$ and $\phi _{ji}$ over the path index $p$, and $\text{(b)}$ employs the identity $\mathbb{E} \left\lbrace \left| \alpha_{ji} \right|^4 \right\rbrace\,{=}\,2\beta_{ji}^2$ and the definition ${\rm E}_{ji}(m)\,{=}\,\mathbb{E} \left\lbrace \exp\left( {-}j 2\pi \frac{D}{\lambda} \left(m\right) \cos\left(\phi_{ji}\right)\right) \right\rbrace$ which is equal to $\mathbf{R}_{ji}^{\phi}(n{+}m,n)$ for any $n\,{\leq}\,M{-}m$ in \eqref{eqn:app:cov_phi_uni}. Note that $\phi_{ij}$ do not have identical distributions with the same parameters over the various subscripts representing the UE and the BS of interest, and we therefore keep the indices in ${\rm E}_{ji}$.
\end{case}

\begin{case}\label{case:case_2}
Assuming $p_1 = p_2, p_3 = p_4$ and $p_1\,{\neq}\,p_3$, the desired expectation ${\rm E}_{\phi}(m,n,{m'},{n'})$ can be given as in \eqref{eqn:app:2nd_moment_e1_6},
where the last line employs the fact that $\phi_{ji,p_1}$ and $\phi_{ji,p_3}$ are uncorrelated for $p_1\,{\neq}\,p_3$.
\begin{figure*}[!htb]
\setcounter{MYtempeqncnt}{\value{equation}}
\setcounter{equation}{41}
\begin{align}
{\rm E}_{\phi}(m,n,{m'},{n'}) &= \dfrac{\beta_{ji}^2}{{N_\mathrm{P}}^2} \sum \limits_{p_1=1}^{N_\mathrm{P}} \sum \limits_{p_3 \neq p_1}^{N_\mathrm{P}} \mathbb{E} \left\lbrace \exp\left( {-}j 2\pi \frac{D}{\lambda} \Big[ (n{-}m)\cos\left(\phi_{ji,p_1}\right){+}({n'}{-}{m'})\cos \left(\phi_{ji,p_3}\right)\Big] \right) \right\rbrace \,, \nonumber \\
&= \dfrac{\beta_{ji}^2}{{N_\mathrm{P}}} (N_\mathrm{P}{-}1) \,{\rm E}_{ji}(n{-}m)\,{\rm E}_{ji}({n'}{-}{m'})\,, \label{eqn:app:2nd_moment_e1_6}
\end{align}
\setcounter{equation}{42}
\hrulefill
\end{figure*}
\end{case}

\begin{case}\label{case:case_3}
Assuming $p_1 = p_4, p_2 = p_3$ and $p_1\,{\neq}\,p_2$, the desired expectation ${\rm E}_{\phi}(m,n,{m'},{n'})$ is obtained by following the derivation steps of Case~\ref{case:case_2} which yields \vspace{-1em}

\small
\begin{align} \label{eqn:app:2nd_moment_e1_7}
{\rm E}_{\phi}(m,n,{m'},{n'}) = \dfrac{\beta_{ji}^2}{{N_\mathrm{P}}} (N_\mathrm{P}{-}1){\rm E}_{ji}({n'}{-}m) {\rm E}_{ji}(n{-}{m'}).
\end{align}
\end{case}\normalsize
Incorporating \eqref{eqn:app:2nd_moment_e1_5}, \eqref{eqn:app:2nd_moment_e1_6}, and \eqref{eqn:app:2nd_moment_e1_7} yields the desired expression of ${\rm E}_{\phi}(m,n,{m'},{n'})$ in \eqref{eqn:2nd_moment_phi}, and ${\rm E}_{11}$ can be computed by employing \eqref{eqn:2nd_moment_phi} in \eqref{eqn:app:2nd_moment_e1_3}.

The second expectation ${\rm E}_{12}$ in \eqref{eqn:app:2nd_moment_e1_2} can be expressed as a weighted sum of scalars as follows
\begin{align}\label{eqn:app:2nd_moment_e1_8}
{\rm E}_{12} & = \sum\limits_{m=1}^{M} \sum\limits_{n=1}^{M} \sum\limits_{m'=1}^{M} \sum\limits_{n'=1}^{M} \tilde{\textbf{R}}_{ii}(m,n) \tilde{\textbf{R}}_{ii}^{\rm H}(m',n') \times
\nonumber \\
& \hspace{12em} \mathbb{E}\left\lbrace h_{ji,n'} h_{ji,m}^{\ast} h_{ki,n} h_{ki,m'}^{\ast} \right\rbrace \,,\nonumber \\
&= \sum\limits_{m=1}^{M} \sum\limits_{n=1}^{M} \sum\limits_{m'=1}^{M} \sum\limits_{n'=1}^{M} \tilde{\textbf{R}}_{ii}(m,n) \tilde{\textbf{R}}_{ii}^{\rm H}(m',n') \times
\nonumber \\
& \hspace{12em} \textbf{R}_{ji}(n',m) \textbf{R}_{ki}(n,m')\,,
\end{align}
where the last line follows from the fact that $k{\neq}j$ as imposed by the summation in \eqref{eqn:app:2nd_moment_e1_2}. Employing \eqref{eqn:app:2nd_moment_e1_3} and \eqref{eqn:app:2nd_moment_e1_8}, we obtain ${\rm E}_{1}$ given in \eqref{eqn:app:2nd_moment_e1_2} as follows
\begin{align}\label{eqn:app:2nd_moment_e1}
{\rm E}_{1} &= \tau^2 \sum\limits_{m=1}^{M} \sum\limits_{n=1}^{M} \sum\limits_{m'=1}^{M} \sum\limits_{n'=1}^{M} \! \tilde{\textbf{R}}_{ii}(m,n) \, \tilde{\textbf{R}}_{ii}^{\rm H}(m',n') \times
\nonumber \\
&\left[ {\rm E}_{\phi}(m,n,m',n') \vphantom{\sum\limits_{\substack{k{=}1; \, k{\neq}j}}^{N_\mathrm{L}} } + \!\!\!\sum\limits_{\substack{k{=}1; \, k{\neq}j}}^{N_\mathrm{L}} \!\!\!\!\textbf{R}_{ji}(n',m) \textbf{R}_{ki}(n,m') \right]\,,
\end{align}
which can be computed by means of ${\rm E}_{\phi}(m,n,m',n')$ given in \eqref{eqn:2nd_moment_phi}. Finally, we consider the expectation ${\rm E}_{2}$ in \eqref{eqn:app:2nd_moment_2}. Defining $\textbf{v}\,{=}\,\textbf{S}\,\tilde{\textbf{R}}_{ii}^{\rm H}\textbf{h}_{ji}$ and $\textbf{C}^{\rm N}\,{=}\,\mathbb{E}\left\lbrace \textbf{n}_{i} \textbf{n}_{i}^{\rm H} \right\rbrace$, ${\rm E}_{2}$ is given as
\begin{align}\label{eqn:app:2nd_moment_e2_1}
{\rm E}_{2} &= \mathbb{E} \left\lbrace \textbf{v}^{\rm H} \textbf{n}_{i} \textbf{n}_{i}^{\rm H} \textbf{v} \right\rbrace
= \sum\limits_{m=1}^{M\tau} \sum\limits_{n=1}^{M\tau} \textbf{C}^{\rm N}(m,n) \mathbb{E}\left\lbrace v_m^{\ast}v_n \right\rbrace
\nonumber \\
&= \sigma^2\sum\limits_{m=1}^{M\tau} \mathbb{E}\left\lbrace \left| v_m \right|^2 \right\rbrace ,
\end{align}
where we employ $\textbf{C}^{\rm N}\,{=}\,\sigma^2 \textbf{I}_{M\tau}$, \color{black} and substituting $\textbf{v}\,{=}\,\textbf{S}\,\tilde{\textbf{R}}_{ii}^{\rm H}\textbf{h}_{ji}$ back in \eqref{eqn:app:2nd_moment_e2_1} yields
\begin{align}\label{eqn:app:2nd_moment_e2_2}
{\rm E}_{2} &= \sigma^2 {\rm tr} \left\lbrace \textbf{S} \tilde{\textbf{R}}_{ii}^{\rm H}\textbf{R}_{ji} \tilde{\textbf{R}}_{ii} \textbf{S}^{\rm H} \right\rbrace .
\end{align}
As a result, substituting \eqref{eqn:app:2nd_moment_e1} and \eqref{eqn:app:2nd_moment_e2_2} in \eqref{eqn:app:2nd_moment_2} yields the desired second order moment in \eqref{eqn:2nd_moment}.

\section{Measure of Channel Hardening}\label{app:harden_measure}
The channel hardening measure $\mathcal{M}_{ij}$ is defined in~\cite{Larsson17ChanHard} as
\begin{align} \label{eqn:app:measure_defn}
\mathcal{M}_{ij} = \frac{{\rm Var} \left\lbrace \left\| \textbf{h}_{ij} \right\|^2 \right\rbrace }{\left( \mathbb{E}  \left\lbrace \left\| \textbf{h}_{ij} \right\|^2 \right\rbrace \right)^2 } = \frac{\mathbb{E} \left\lbrace \left\| \textbf{h}_{ij} \right\|^4 \right\rbrace   }{ \left( \mathbb{E}  \left\lbrace \left\| \textbf{h}_{ij} \right\|^2 \right\rbrace \right)^2 } - 1\,,
\end{align}
with $\mathbb{E} \left\lbrace \left\| \textbf{h}_{ij} \right\|^2 \right\rbrace\,{=}\,\textrm{tr} \lbrace \mathbf{R}_{ij} \rbrace\,{=}\,\beta_{ij} M$ second order moment, and the fourth order moment $\mathbb{E} \left\lbrace \left\| \textbf{h}_{ij} \right\|^4 \right\rbrace$ given as
\begin{align*}
\mathbb{E} \left\lbrace \left\| \textbf{h}_{ij} \right\|^4 \right\rbrace &= \frac{1}{{N_\mathrm{P}}^2} \sum\limits_{p_1=1}^{N_\mathrm{P}} \sum\limits_{p_2=1}^{N_\mathrm{P}} \sum\limits_{p_3=1}^{N_\mathrm{P}}  \sum \limits_{p_4=1}^{N_\mathrm{P}} \mathbb{E}_{\alpha} \times
\nonumber \\
&\mathbb{E} \left\lbrace \textbf{a}^{\rm H}\!\left( \phi_{ij, p_1}\right)  \textbf{a}\left( \phi_{ij, p_2}\right) \textbf{a}^{\rm H}\!\left( \phi_{ij, p_3}\right) \textbf{a}\left( \phi_{ij, p_4}\right) \right\rbrace ,
\end{align*}
with $\mathbb{E}_{\alpha}\,{=}\,\mathbb{E} \left\lbrace  \alpha_{ij, p_1}^{\ast} \alpha_{ij, p_2} \alpha_{ij, p_3}^{\ast} \alpha_{ij, p_4} \right\rbrace$. Similar to the discussion for \eqref{eqn:app:2nd_moment_e1_4}, $\mathbb{E}_{\alpha}$ is nonzero only when 1) $p_1\,{=}\,p_2\,{=}\,p_3\,{=}\,p_4$, 2) $p_1\,{=}\,p_2$, $p_3\,{=}\,p_4$, $p_1\,{\neq}\,p_3$ or 3) $p_1\,{=}\,p_4$, $p_2\,{=}\,p_3$, $p_1\,{\neq}\,p_2$ and hence \vspace{-2em}

\small \begin{align*}
&\mathbb{E} \left\lbrace \left\| \textbf{h}_{ij} \right\|^4 \right\rbrace =  \frac{ \left(1 + N_\mathrm{P} \right) \beta_{ij}^2 M^2}{{N_\mathrm{P}}}
\nonumber \\
& +  \frac{\beta_{ij}^2}{{N_\mathrm{P}}^2} \sum \limits_{p_1=1}^{N_\mathrm{P}} \sum \limits_{p_2{\neq}p_1}^{N_\mathrm{P}} \mathbb{E} \left\lbrace \textbf{a}^{\rm H}\!\left( \phi_{ij, p_1}\right)  \textbf{a}\left( \phi_{ij, p_2}\right) \textbf{a}^{\rm H}\!\left( \phi_{ij, p_2}\right)  \textbf{a}\left( \phi_{ij, p_1}\right) \right\rbrace \,,
\end{align*} \normalsize
where $\mathbb{E} \left\lbrace \left| \alpha_{ij} \right|^4 \right\rbrace\,{=}\,2\beta_{ij}^2$ and $ \textbf{a}^{\rm H}\!\left( \phi_{ij, p}\right) \textbf{a}\left( \phi_{ij, p}\right)\,{=}\,M$. Substituting the second and the fourth order moments in \eqref{eqn:app:measure_defn}, the channel hardening measure can be given as in \eqref{eqn:app:measure_hardening}.
\begin{figure*}
\setcounter{MYtempeqncnt}{\value{equation}}
\setcounter{equation}{48}
\begin{align} \label{eqn:app:measure_hardening}
\mathcal{M}_{ij} &= \frac{1}{N_\mathrm{P}} {+} \frac{1}{N_\mathrm{P}^2 M^2} \sum\limits_{p_1=1}^{N_\mathrm{P}} \sum \limits_{p_2{\neq}p_1}^{N_\mathrm{P}} \mathbb{E} \left\lbrace \textbf{a}^{\rm H}\!\left( \phi_{ij, p_1}\right)  \textbf{a}\left( \phi_{ij, p_2}\right)\textbf{a}^{\rm H}\!\left( \phi_{ij, p_2}\right)\textbf{a}\left( \phi_{ij, p_1}\right) \right\rbrace \,, \nonumber \\
&= \frac{1}{N_\mathrm{P}} {+} \frac{1}{N_\mathrm{P}^2 M^2} \sum \limits_{p_1=1}^{N_\mathrm{P}} \sum \limits_{p_2{\neq}p_1}^{N_\mathrm{P}}  \sum \limits_{n=1}^{M} \sum \limits_{m=1}^{M}   \mathbb{E} \left\lbrace \exp\left( {-}j 2\pi \frac{D}{\lambda} \Big[ (n{-}m) \left( \cos\left(\phi_{ij,p_2}\right){-}\cos\left(\phi_{ij,p_1}\right) \right) \Big] \right) \right\rbrace
\end{align}
\setcounter{equation}{49}
\hrulefill
\end{figure*}
Taking the terms for the equality of $m = n$ out of the summation, and employing the angular covariance matrix $\mathbf{R}_{ij}^{\phi}$ given in \eqref{eqn:app:cov_phi_element}, we end up with \vspace{-1em}

\small
\begin{align} \label{eqn:app:measure_last}
&\mathcal{M}_{ij} = \frac{1}{N_\mathrm{P}} {+} \frac{N_\mathrm{P}-1}{N_\mathrm{P} M}
\nonumber \\
&\quad\; {+} \frac{1}{N_\mathrm{P}^2 M^2} \sum \limits_{p_1=1}^{N_\mathrm{P}} \sum \limits_{p_2{\neq}p_1}^{N_\mathrm{P}}  \sum \limits_{n=1}^{M} \sum \limits_{m{\neq}n}^{M} \mathbf{R}_{ij}^{\phi}(n,m) \mathbf{R}_{ij}^{\phi}(m,n)\,, \nonumber\\
&= \frac{1}{N_\mathrm{P}} {+} \frac{N_\mathrm{P}-1}{N_\mathrm{P} M}
\nonumber\\
&{+} \left( \frac{1}{N_\mathrm{P}^2 M^2} \left( N_\mathrm{P}-1 \right) N_\mathrm{P}  \sum \limits_{n=1}^{M} \sum \limits_{m{\neq}n}^{M} \mathbf{R}_{ij}^{\phi}(n,m) \mathbf{R}_{ij}^{\phi}(m,n)\,\right) , \\
&= \frac{1}{N_\mathrm{P}} {+} \frac{N_\mathrm{P}-1}{N_\mathrm{P} M} {+} \frac{N_\mathrm{P}-1}{N_\mathrm{P} M^2} \sum \limits_{n=1}^{M} \sum \limits_{m{\neq}n}^{M} \mathbf{R}_{ij}^{\phi}(n,m) \mathbf{R}_{ij}^{\phi}(m,n) \,, \end{align} \normalsize
where (50) is due to the fact that $\mathbf{R}_{ij}^{\phi}(n,m)$ does not dependent on the path index. Let us consider the term $\sum \limits_{n=1}^{M} \sum \limits_{m{\neq}n}^{M} \mathbf{R}_{ij}^{\phi}(n,m) \mathbf{R}_{ij}^{\phi}(m,n)$. We can represent it as,

\small
\begin{align} \label{eqn:hardening_7}
&\sum \limits_{n=1}^{M} \sum \limits_{m{\neq}n}^{M} \mathbf{R}_{ij}^{\phi}(n,m) \mathbf{R}_{ij}^{\phi}(m,n) = \sum \limits_{n=1}^{M} \sum \limits_{m{=}1}^{M} \mathbf{R}_{ij}^{\phi}(n,m) \mathbf{R}_{ij}^{\phi}(m,n)
\nonumber \\
&\hspace{15em} - \sum \limits_{m=1}^{M} \left| \mathbf{R}_{ij}^{\phi}(m,m) \right|^2
\nonumber \\
&\hspace{3em}= \sum \limits_{n=1}^{M} \sum \limits_{m{=}1}^{M} \left|\mathbf{R}_{ij}^{\phi}(m,n) \right|^2 - \sum \limits_{m=1}^{M} \left| \mathbf{R}_{ij}^{\phi}(m,m) \right|^2.
\end{align} \normalsize  Due to the conjugate symmetry of the covariance matrix, $\left| \mathbf{R}_{ij}^{\phi}(m,n) \right|^2 = \left| \mathbf{R}_{ij}^{\phi}(n,m) \right|^2$. By making use of this fact \eqref{eqn:hardening_7} can be simplified and given as, \begin{align} \label{eqn:hardening_8}
\sum \limits_{n=1}^{M} \sum \limits_{m{\neq}n}^{M} \mathbf{R}_{ij}^{\phi}(n,m) \mathbf{R}_{ij}^{\phi}(m,n) =\sum \limits_{m=1}^{M-1} \sum \limits_{n{=}1}^{M-m} 2 \left|\mathbf{R}_{ij}^{\phi}(m+n,n) \right|^2.
\end{align} Employing ${\rm E}_{ij}(m)\,{=}\,\mathbf{R}_{ij}^{\phi}(n{+}m,n)$ for any $n\,{\leq}\,M{-}m$, \eqref{eqn:hardening_8} can be represented as \begin{align} \label{eqn:hardening_9}
\sum \limits_{n=1}^{M} \sum \limits_{m{\neq}n}^{M} \mathbf{R}_{ij}^{\phi}(n,m) \mathbf{R}_{ij}^{\phi}(m,n) = \sum \limits_{m=1}^{M-1}  2 (M-m)\left|{\rm E}_{ij}(m) \right|^2.
\end{align}  Using \eqref{eqn:hardening_9} in (51) we obtain the channel hardening measure in \eqref{eqn:harden_measure}. \color{black}

\bibliographystyle{IEEEtran}
\bibliography{papers}

\end{document}